\documentclass[1p, 12pt]{elsarticle}

\usepackage{lineno}

\usepackage{hyperref}
\hypersetup{colorlinks  = true}

\journal{Respiratory Physiology \& Neurobiology}
\biboptions{authoryear}
\usepackage{amsmath}
\usepackage{caption}
\usepackage{subcaption}
\usepackage{graphicx}
\usepackage{upgreek}
\graphicspath{ {img/} }
\usepackage{geometry}
\usepackage{array}

\usepackage{setspace}
 
\usepackage{upgreek} 

\let\today\relax
\makeatletter
\def\ps@pprintTitle{%
    \let\@oddhead\@empty
    \let\@evenhead\@empty
    \def\@oddfoot{\footnotesize\itshape
         {Accepted in: Respiratory Physiology \& Neurobiology (doi.org/10.1016/j.resp.2021.103719)} \hfill\today}%
    \let\@evenfoot\@oddfoot
    }
\makeatother

\begin{document}

\begin{frontmatter}
	\title{The impact of nasal adhesions on airflow and mucosal cooling - a computational fluid dynamics analysis}
	\author[label1]{Praween Senanayake}
	\author[label2]{Hana Salati}
	\author[label3]{Eugene Wong}
	\author[label1]{Kimberley Bradshaw}
	\author[label3]{Yidan Shang}
	\author[label1,label4]{Narinder Singh}
	\author[label3]{Kiao Inthavong \corref{cor1}}
	
	\cortext[cor1]{Corresponding Author: kiao.inthavong@rmit.edu.au}
	\address[label1]{Department of Otolaryngology, Head and Neck Surgery, Westmead Hospital, Sydney, NSW, Australia}
	\address[label3]{Mechanical \& Automotive Engineering, School of Engineering, RMIT University, Bundoora, Victoria 3083, Australia}
	\address[label2]{Auckland University of Technology}
	\address[label4]{Faculty of Medicine \& Health, The University of Sydney, NSW 2006, Australia}

{\setstretch{1.0}	
\begin{abstract}
Nasal adhesions are a known postoperative complication following surgical procedures for nasal airway obstruction (NAO); and are a common cause of surgical failure, with patients often reporting significant NAO, despite relatively minor adhesion size. Division of such nasal adhesions often provides much greater relief than anticipated, based on the minimal reduction in cross-sectional area associated with the adhesion. The available literature regarding nasal adhesions provides little evidence examining their quantitative and qualitative effects on nasal airflow using objective measures. This study examined the impact of nasal adhesions at various anatomical sites on nasal airflow and mucosal cooling using computational fluid dynamics (CFD). A high-resolution CT scan of the paranasal sinuses of a 25-year-old, healthy female patient was segmented to create a three-dimensional nasal airway model. Virtual nasal adhesions of 2.5~mm diameter were added to various locations within the nasal cavity, representing common sites seen following NAO surgery. A series of models with single adhesions were created. CFD analysis was performed on each model and compared with a baseline no-adhesion model, comparing airflow and heat and mass transfer. The nasal adhesions resulted in no significant change in bulk airflow patterns through the nasal cavity. However, significant changes were observed in local airflow and mucosal cooling around and immediately downstream to the nasal adhesions. These were most evident with anterior nasal adhesions at the internal valve and anterior inferior turbinate. Postoperative nasal adhesions create local airflow disruption, resulting in reduced local mucosal cooling on critical surfaces, explaining the exaggerated perception of nasal obstruction. In particular, anteriorly located adhesions created greater disruption to local airflow and mucosal cooling, explaining their associated greater subjective sensation of obstruction.
\end{abstract}
}	

\footnotesize{Accepted in: Respiratory Physiology \& Neurobiology (doi.org/10.1016/j.resp.2021.103719)}

\end{frontmatter}

\section{Introduction}
\label{Introduction}
One of the most common complications arising from nasal airway obstruction surgery is the formation of nasal adhesions, also known as synechiae or scar bands. Nasal adhesions are abnormal tissue bridges between adjacent mucosal surfaces, particularly between the septum and components of the lateral nasal wall, such as the inferior and middle turbinates. The adhesions most commonly form post-operatively following sinonasal surgery or can form following trauma. Less common causes of adhesions include chronic inflammatory conditions such as granulomatosis with polyangiitis (Wegener’s granulomatosis), sarcoidosis, systemic lupus erythematosus, tuberculosis or following radiotherapy \citep{Vartiainen1992, Jones1999}. 

Nasal adhesions are thought to form when two opposing raw or inflamed mucosal surfaces heal inappropriately, forming a mucosal bridge \citep{Choi2017}. The location of post-operative nasal adhesions typically depends primarily upon specific patient anatomy and the area of surgical intervention \citep{Nayak1998}. The adhesions are most commonly found at the internal nasal valve following septoplasty or the middle meatus following endoscopic sinus surgery but can occur at other locations \citep{Chandra2009}. The reported incidence of adhesions following septoplasty with submucous turbinate surgery estimates 0-19.7\% occurrence \citep{Naghibzadeh2011, Deniz2014, Chen2019, Joshi2019}.

Patients who develop nasal adhesions after surgery often complain of persistent nasal airway obstruction \citep{Young1997, Stewart2004, JoeJacob2011, Derin2016}. In particular, the degree of perceived nasal airway obstruction that patients report following nasal adhesion formation appears out-of-proportion compared to the relatively small size of the adhesion and minimal reduction in the cross-sectional area typically seen \citep{Becker2010}. Furthermore, the division of such small adhesions is often accompanied by significantly greater relief of nasal airway obstruction than would be anticipated from what appears to be a relatively minor increase in cross-sectional area \citep{Henriquez2013}. More anterior adhesions appear to have an even greater impact on perceived nasal airway obstruction than posterior adhesions \citep{Rettinger2006}. In anterior adhesions, the effect on cross-sectional area is at least partly explained by the relatively narrower dimensions at the internal nasal valve. Despite the observed significant impact of post-operative adhesions, there remains a lack of objective evidence in the literature that examines the cause for this disproportionate symptom profile, nor the physiological mechanism by which this occurs. To the best of our knowledge, there have been no studies assessing nasal adhesions with traditional objective measures of nasal airflow, such as rhinomanometry, acoustic rhinometry or peak nasal inspiratory flow (PNIF).

In recent years, the development of Computational Fluid Dynamics (CFD) has facilitated reliable assessment of airflow within the sinonasal cavity. In contrast to traditional objective measures of airflow, CFD provides highly detailed, reproducible and quantifiable results, where multiple variables can be rapidly assessed while avoiding many of the challenges of physical experiments. CFD has been validated and utilised in the assessment of nasal airflow in various physiological conditions, including septal deviation, sinus ventilation, nasal cycling, and examining the efficacy of nasal irrigation and sprays \citep{Lindemann2013, Nishijima2018, Alam2019, FrankIto2019, Abouali2012, Garcia2015, Tian2017, Inthavong2020a, Dong2018, Kim2014, Salati2021, Gaberino2017, Zhu2012, Jo2015}.

This study used CFD to examine the effect of nasal adhesions on nasal airflow in locations commonly seen following NAO surgery (i.e. septoplasty and turbinate reduction). We hypothesise that the disproportionally excessive NAO observed with postoperative nasal adhesions may be more related to key alterations in local nasal airflow and mucosal cooling downstream to the adhesion, rather than due to the relatively minor reduction in cross-sectional area. Furthermore, we hypothesise that anterior nasal adhesions create greater local disruption in airflow in the critical internal valve region and that reduction in cross-sectional area may have a proportionally greater impact in this already narrow region.

\section{Method}
\subsection{Normal nasal airway model}
A high-resolution Computed Tomography (CT) scan of a healthy 25-year-old female patient was used to create the nasal airway computational model. The patient had no history of previous sinonasal pathology, trauma or surgery and no anatomical abnormalities. CT scanning of the nasal airway was conducted using a Siemens Dual Source CT Scanner (Siemens Healthcare, Erlangen, Germany) with the following imaging parameters: $0.39 \times 0.39 $mm pixel size, $512 \times 512$ pixel image dimensions and slice thickness of 0.6 mm. Written informed consent to participate in the study was obtained from the patient before commencement. IRB approval was obtained (Western Sydney Local Health District, Human Research Ethics Committee; Approval number: ETH08671). 3D Slicer segmentation software was used to create a three-dimensional model of the nasal airway from the CT scan. As previous studies have demonstrated that the paranasal sinuses do not contribute significantly to nasal airflow \citep{Xiong2008, Ge2012}, the frontal, maxillary, ethmoid and sphenoid sinuses were removed from the model. The geometry of the face was retained, and an enclosed hemisphere representing the outer surrounding air was constructed in front of the face, which was set as a pressure inlet condition (Fig \ref{fig:location}A).  

\subsection{Nasal adhesion models}
The reference model with no adhesions was edited, and a further five nasal adhesion models were created separately by placing solid transverse cylinders in various locations within the right side of the nasal cavity using ANSYS Spaceclaim\textregistered (Ver 20R1). The cylinder diameters were 2.5 mm, and their locations were consistent with postoperative adhesions commonly seen following NAO surgery (i.e. septoplasty and turbinate reduction). The locations do not represent common sites following endoscopic sinus surgery, which will be investigated in a future study. Fig \ref{fig:location}B demonstrates the adhesion sites chosen; however, each individual model only contained one single adhesion. All virtual nasal adhesions were added under a tertiary fellowship-trained rhinologist (NS) guidance to ensure the models were consistent and clinically accurate. In clinical practice, nasal adhesions are often accompanied by medialisation of the inferior turbinate or middle turbinate. For this study, we chose not to model such medialisation to analyse the effect attributable to the adhesions alone.

The no adhesion model (control) and adhesion models were exported as Standard Tessellation Language (STL) files, and ANSYS Fluent\textregistered (Ver 20R1) was used for mesh generation. Our mesh strategy was based on selecting the recommended  mesh element size functions evaluated in \cite{Inthavong2018} where a curvature size of 0.45~mm was found to be sufficient for mesh independence in a nasal cavity. A much larger mesh size function (8~mm) was defined at the hemispherical outer dome for mesh optimisation. Five prism layers were attached to all wall boundaries to ensure the boundary layer was captured. A poly-hexcore mesh was created, which consisted of polyhedral surface elements and hexahedral elements in the internal cavity space. The poly-hexcore mesh was used because it produces significantly fewer elements to tetrahedral meshing, approximately 3.5$\times$ reduction while using the same size functions. We started with a mesh size function of 0.45~mm, which produced a total of 1.7~million elements (equivalent to 5.3~million tetrahedral cells). A second refined mesh was created by decreasing the size function to 0.35mm which produced 1.84~million poly-hexcore cells (equivalent to 8.4~million tetrahedral cells). Comparatively, \cite{frank2016influence, Inthavong2018} both reported a mesh independent nasal model with 4~million tetrahedral cells. A mesh independence analysis of temperature and humidity profiles at two coronal cross-sections is given in the Appendix / Supplementary Material (Supplementary 1). Using the same mesh size functions, the number of poly-hexcore  mesh elements for the six models ranged from 1.42 to 1.87 million cells. 

\subsection{Fluid flow simulation}
A constant inhalation flow rate of 15 L/min was applied, and a laminar, incompressible flow model was used. A laminar flow regime was assumed to exist throughout the cavity at the same constant flow rate of 15 L/min. Several studies have reported that under the flow rate up to 15 L/min, the nasal airflow can be considered quasi-steady \citep{Doorly2008,Burgos2017}. Therefore, it was assumed that the laminar flow assumption does not cause a significant error \citep{Na2020, shang2019}. The second-order upwind scheme was used for the spatial discretization of the governing equations, and the coupled-scheme was used for the pressure-velocity coupling. The flow equations describing the conservation for mass, momentum,  energy, and  transport for the mass fraction of water vapour are expressed as:
\begin{align}
\rho \{\frac{\partial u}{\partial t} + ({u} \cdot \nabla){u}\} &= -\nabla {p} + \mu \nabla ^2 {u} 
\end{align}
  \begin{align}
\nabla \cdot {u} &= 0
\end{align}
\begin{align}
\rho {C_p} \{\frac{\partial T}{\partial t} + ({u} \cdot \nabla){T}\} &= {K}\nabla ^2 {T}
\end{align}
\begin{align}
\frac{\partial F}{\partial t} + ({u} \cdot \nabla) {F} &= {D}\nabla ^2 {F}
\end{align}
where $\rho$, $t$, $u$, $p$, $mu$, $C_p$, $T$, and $K$ are density, time, velocity, pressure, viscosity, specific heat capacity, temperature , and thermal conductivity. In Equation 4, $F$ and $D$ denote mass fraction of water vapour and mass diffusion coefficient, respectively. The simulation was considered to be steady-state, and it was assumed that a converged solution was reached when the residuals reached below 10 e$^{-5}$ except for energy which was taken to 10 e$^{-6}$. The inlet temperature at the outer dome boundary was 20$^\circ$C and 35\% relative humidity.

\subsection{Submucus wall model}
The wall boundary was set with a virtual thickness with length $L$ to represent the mucus and submucus layers (see Supplementary Material 2). An effective thermal resistance,  $R = L/kA$ was applied where $A$ is the area normal to the conduction direction and was taken as per unit area. The remaining conductivity and thickness parameters were defined following \citep{Na2020} who reviewed anatomic information on the respiratory mucosa from \citep{Beule2010}. The effective resistance used was  $R_{eff} = 0.020$K/W.

Following measurements of patient data that included body temperature, the temperature along the bottom of the wall model surface was set to $36^\circ$ in all nasal cavity surfaces, except for the nasal vestibule, which was set to $34^\circ$, slightly less based on its external location to the internal nasal cavity. The mucosal wall was assumed to be wet and have a saturated state, and its water-vapour mass fraction is defined at the surface boundary with a 100\% relative humidity condition. Since the surface boundary temperature varies with the flow conditions, an approximate function relating the saturated water-vapour concentration with temperature was created, giving the following equation:
\begin{equation}
	C_{\mathrm{surface}} =  \frac{1}{1000}(0.0006312T^3-0.01097^2+0.6036T+2.027)
\end{equation}
where $C_{\mathrm{surface}}$ is the water vapour concentration in kg.s$^{-1}$ and $T$ is the temperature in $^\circ$C.

\section{Results}
\subsection{Flow streamlines}
The effect of adhesions on nasal airflow was explored through streamlines in the right lateral view shown in Fig \ref{fig:streamlines}. The baseline 'no adhesion' model demonstrated initial oblique high-velocity flow at the nasal vestibule, followed by a decrease in airflow velocity occurring in the middle meatus region, and further decreasing in the curvature down to the nasopharynx. All models demonstrated flow re-circulation in the nasal vestibule which corresponds to the scroll region between the upper and lower lateral cartilages \citep{Inthavong2019, Zhao2004, Ito2017, lu2018part}. This flow circulation was diminished in the internal nasal valve (IV) adhesion model compared to baseline, where the adhesion redirected part of the airflow toward the superior vestibule. In the anterior inferior turbinate (AIT) model, there was a pronounced vertical trajectory in the streamlines secondary to the adhesion, with airflow diverted from the inferior meatus and concentrated in the middle meatus. 

The resistance caused by the nasal adhesions in both the middle inferior turbinate (MIT) and posterior inferior turbinate (PIT) models resulted in diversion of airflow to the middle meatus and medial to the middle turbinate, with a corresponding reduction at the inferior meatus. In contrast, the middle turbinate (MT) model showed flow separation around the adhesion, resulting in reduced middle meatal flow and disturbed streamlines in the posterior, inferior region.

\subsection{Velocity, temperature, and relative humidity in coronal planes}
Velocity contours at coronal slices for all models are given in Fig \ref{fig:vel-conts}. Generally, the flow upstream of the adhesion does not differ in comparison with the baseline no-Adhesion model. The presence of an adhesion separates the flow into a superior and inferior flow region, but this is only relevant in the bulk flow region where high velocities are displaced. Where the velocity is low, the effects of the adhesion is minimal. After each adhesion, the downstream flow recovers and is similar to the velocity contours in the baseline model.

Fig \ref{fig:temp-conts} depicts the temperature contours at coronal slices for all models. The inlet air temperature was 20$^\circ$C, and the cooler temperature locations correspond with the high-velocity locations (found in Fig \ref{fig:vel-conts}). In general, the adhesion disturbs the temperature field in the vicinity of the adhesion, and the effects diminish with distance moving downstream.

Fig \ref{fig:rh-conts} depicts the relative humidity contours at coronal slices for all models where the inhaled air is drier and enters the nostrils at 35\%. It influences the inferior half of the coronal planes at slices 1 and 2 before the flow changes. From slice-3 onward, the mass transfer from the wall boundary provides rapid diffusion into the airflow. The presence of the adhesions is most influential at the IV adhesion.

The velocity, temperature, and relative humidity contours were mass-weighted-averaged for the left and right cavities of the no-adhesion model, providing a single averaged value at each coronal slice location. These were plotted in Fig \ref{fig:noadh} and thus provided a quantitative baseline representing the adhesion models compared to the adhesion models. The averaged velocity magnitude was highest at the anterior half and progressively decreased as the cross-sectional area increased posteriorly. Both the temperature and relative humidity increased steadily from the inlet conditions (20$^\circ$C and 35\% relative humidity)
towards the wall conditions of 36$^\circ$C and 100\%.

Similarly, the velocity, temperature, and relative humidity were averaged on each coronal slice for the five adhesion models. The effect of an adhesion was then quantified by taking a percentage difference from the baseline values as:
\[
\frac{\phi_{\textup{adhesion}} -\phi_{\textup{noAdhesion}}}{\phi_{\textup{noAdhesion}}}
\]
Figure \ref{fig:cont-difference} showed negligible change in the left cavity that had no adhesion. In the right cavity, adhesions altered all three flow parameters, velocity, temperature, and relative humidity. The change was most significant for adhesions located most anteriorly. In order of anterior to posterior adhesion, this was IV, AIT, MT, MIT, followed by PT.  

The surface heat flux in summarised in Figure \ref{fig:heatFlux} (and given in detail in Supplementary material 4) showed the left cavity was unaffected by the nasal adhesion in the right cavity. The nasal adhesion affected the surface heat flux locally around the location of the adhesion compared to the No-Adhesion model and was most significant for the IV model, where the surface heat flux varied from 2.2 W/m$^2$ (no-adhesion) to 15.6 W/m$^2$. As the air temperature increased from the nostrils to the nasopharynx (Figure \ref{fig:temp-conts}), the surface heat flux decreased for all models. The surface heat flux at the adhesion plane for each model was higher than the No-Adhesion model. The presence of the nasal adhesion increased the air residence and exposure time with the nasal mucosal surface locally which enhanced the heat transfer between the mucosal wall and inhaled air. 

The influence of an adhesion was determined by taking the local pressure differential between the cross-section planes prior to and after the adhesion location given by
\[
\Delta P_n = \mathrm{abs}(P_{n+1} - P_{n-1})
\]
and these values were plotted in Fig \ref{fig:press}. The largest pressure differential occurred at the internal valve (IV) region, and the presence of an adhesion increased the differential by a further $42\%$. The smallest pressure differential occurred at the middle inferior turbinate (MIT) region, where there is low flow velocity. The presence of an adhesion increases the pressure differential by more than other regions as a percentage the no-Adhesion control model. 

The effect of adhesions on nasal mucosal wall surface temperature for the right nasal cavity, including the medial septal surface and lateral nasal wall, is given in Fig\ref{fig:surfSubt}. The No-Adhesion model demonstrated the heat transfer mechanism of the nasal cavity where the bulk airflow containing cool air entered the nose and cooled the surfaces it came into contact with. As the flow moved downstream, heat exchange took place rapidly and the air was warmed by the nasal mucosa. All adhesion models showed a similar transition, however, a clear area of increased mucosal temperature was evident in the wake of each adhesion. This was particularly significant in the anterior nasal adhesions (IV and AIT) on the septal surface and, to a lesser extent, on the lateral nasal wall surface.

\section{Discussion}
The presence of adhesions in different locations of the nasal cavity lead to different effects on the flow behaviour in the nasal cavity. The results showed that anteriorly located adhesions were more influential in altering the flow field, and the heat transfer, with the anterior-most adhesion being the internal nasal valve which exhibited the most significant changes in all flow parameters. It is established that the anterior portion of the nasal cavity is responsible for up to 80\% of the air conditioning capacity of the respiratory airway \citep{keck2000, elad2008air,Burgos2017}, and high resistance \citep{zhu2011eval,van2021pressure}. As a result, disturbances that impede the natural flow of air in this region create extremely sensitive responses to the air conditioning and resistance that are likely to cause perceptions of airway obstruction. Recent findings corroborate, with anterior adhesions appearing to significantly impact perceived nasal airway obstruction than posterior adhesions (Rettinger and Kirsche 2006). 

The nasal adhesion were 2.5mm diameter cylinders in size that represented adhesions commonly seen in clinical practice. The inclusion of adhesions had little to no change in overall resistance or bulk airflow through the nasal cavity because the flow field recovers further downstream of the adhesion. However, the effects were more localised with changes around the adhesion itself and whether it impeded the bulk flow regions. In the anterior nasal cavity, where the cross-sections are small, the flow field is easily disturbed even by small adhesions. This was most evident by the IV and AIT adhesions, which correspond to the greater symptoms observed with adhesions in these regions in clinical practice \citep{Wang2016}. 

Adhesions within the nasal cavity altered the flow locally, which generally decreased the temperature in the surrounding regions. For instance, the redirected flow in the IV model moved superiorly in the nasal valve region, which cooled down the mucosal surface compared to the no-adhesion model. \cite{SINGHA2010757} suggested the trailing wake behind a  cylinder becomes more unstable with higher velocity, and the reattachment of the separated flow occurs at a further distance. Hence, the mucosal cooling differences between the adhesion models, which were placed in high-velocity regions (IV and AIT) and no adhesion model, are more notable than the other adhesion models.

In the AIT model, the redirected flow over the adhesion caused a lower temperature on the sides of the adhesion and reduced the septal surface temperature compared to the no-Adhesion model. The lowest temperature difference around the adhesion cylinder was found for the PIT model. The PIT adhesion did not intercept the bulk airflow, resulting in a minor change in all flow variables. The adhesion in these models affected the septal surface temperature more than the lateral walls because the bulk fluid flow mainly remains close to the septal wall. 

The finding that there is no significant reduction in overall bulk nasal airflow at first glance appears to contradict patient-reported complaints of the degree of perceived NAO associated with postoperative nasal adhesion. This contradiction may be resolved through application of the mucosal cooling theory of nasal patency, as identified by both \cite{Zhao2011} and \cite{Sullivan2014} who found that mucosal cooling was the most important predictor of patient perception of nasal patency, rather than direct detection of nasal airflow. The results of our study support this theory, where adhesions found anteriorly in the nasal cavity have significant alterations to the temperature and humidity field. While the airflow temperatures in the cavity did not change significantly, the surface temperature contours on the mucosal wall showed reduced mucosal cooling in the localised areas immediately downstream to the nasal adhesion. These areas may coincide with areas rich in transient receptor potential melastatin family member 8 (TRPM8) thermoreceptors, in which case this localised reduction in mucosal cooling could likely be misinterpreted centrally as an exaggerated degree of nasal airway obstruction. \cite{Keh2011} and \cite{Liu2015} recently used immunohistochemistry to show that high densities of TRPM8 receptors reside in the sub-epithelial layers of the nasal mucosa, with profuse fibres surrounding blood vessels in deeper regions. 

Interestingly, the IV model in the present study showed marked disruption to mucosal cooling in the region of Kisselbach’s plexus in the nasal septum, an area known for its marked vascularity. Disruption of mucosal cooling in this crucial area may contribute to the seemingly exaggerated obstructive symptoms seen in patients with anteriorly located adhesion. Moreover, a recent CFD study by \cite{Zhao2014} identified the region immediately posterior to the nasal vestibule as the site of peak heat loss in respiration as well as being significantly correlated with perceived nasal patency in 44 healthy volunteers. This may help explain why anterior nasal adhesions have been associated with the greatest degree of subjective complaint. In addition, our study shows redistribution of airflow streamlines away from critical surfaces implicated in regulating and assessing inspired air such as the middle and inferior turbinates. The AIT and MT models, in particular, showed marked redirection of local airflow, which coincided with reduced mucosal cooling and would likely interfere with the perception of nasal patency.

\section{Conclusion}
This study used CFD analysis on a single patient with virtual nasal adhesions, located at the sites commonly seen in clinical practice following nasal airway surgery (septoplasty with turbinate reduction), to demonstrate that the presence of postoperative nasal adhesions results in no significant change in overall airflow but does result in localised downstream disruption to airflow. This localised disruption creates reduced local mucosal cooling on critical surfaces, resulting in the exaggerated perception of nasal obstruction, in excess of the minimal reduction in cross-sectional area seen. Adhesions located anteriorly create greater local disruption to airflow and mucosal cooling in critical areas, explaining their associated greater subjective sensation of obstruction.

\section{Acknowledgements}
The authors gratefully acknowledge the financial support provided by the Garnett Passe and Rodney Williams Foundation Conjoint Grant 2019-22.

\small
{\setstretch{1.0}
\bibliography{adhesion}

\begin{thebibliography}{54}
\expandafter\ifx\csname natexlab\endcsname\relax\def\natexlab#1{#1}\fi
\providecommand{\url}[1]{\texttt{#1}}
\providecommand{\href}[2]{#2}
\providecommand{\path}[1]{#1}
\providecommand{\DOIprefix}{doi:}
\providecommand{\ArXivprefix}{arXiv:}
\providecommand{\URLprefix}{URL: }
\providecommand{\Pubmedprefix}{pmid:}
\providecommand{\doi}[1]{\href{http://dx.doi.org/#1}{\path{#1}}}
\providecommand{\Pubmed}[1]{\href{pmid:#1}{\path{#1}}}
\providecommand{\bibinfo}[2]{#2}
\ifx\xfnm\relax \def\xfnm[#1]{\unskip,\space#1}\fi
\bibitem[{Abouali et~al.(2012)Abouali, Keshavarzian, {Farhadi Ghalati},
  Faramarzi, Ahmadi and Bagheri}]{Abouali2012}
\bibinfo{author}{Abouali, O.}, \bibinfo{author}{Keshavarzian, E.},
  \bibinfo{author}{{Farhadi Ghalati}, P.}, \bibinfo{author}{Faramarzi, A.},
  \bibinfo{author}{Ahmadi, G.}, \bibinfo{author}{Bagheri, M.H.},
  \bibinfo{year}{2012}.
\newblock \bibinfo{title}{{Micro and nanoparticle deposition in human nasal
  passage pre and post virtual maxillary sinus endoscopic surgery}}.
\newblock \bibinfo{journal}{Respiratory Physiology and Neurobiology}
  \DOIprefix\doi{10.1016/j.resp.2012.03.002}.
\bibitem[{Alam et~al.(2019)Alam, Li, Bradburn, Zhao and Lee}]{Alam2019}
\bibinfo{author}{Alam, S.}, \bibinfo{author}{Li, C.},
  \bibinfo{author}{Bradburn, K.H.}, \bibinfo{author}{Zhao, K.},
  \bibinfo{author}{Lee, T.S.}, \bibinfo{year}{2019}.
\newblock \bibinfo{title}{Impact of middle turbinectomy on airflow to the
  olfactory cleft: A computational fluid dynamics study}.
\newblock \bibinfo{journal}{Am J Rhinol Allergy} \bibinfo{volume}{33},
  \bibinfo{pages}{263--268}.
\newblock \URLprefix \url{https://www.ncbi.nlm.nih.gov/pubmed/30543120},
  \DOIprefix\doi{10.1177/1945892418816841}.
\bibitem[{Becker et~al.(2010)Becker, Ransom, Guy and Bloom}]{Becker2010}
\bibinfo{author}{Becker, D.G.}, \bibinfo{author}{Ransom, E.},
  \bibinfo{author}{Guy, C.}, \bibinfo{author}{Bloom, J.}, \bibinfo{year}{2010}.
\newblock \bibinfo{title}{Surgical treatment of nasal obstruction in
  rhinoplasty}.
\newblock \bibinfo{journal}{Aesthet Surg J} \bibinfo{volume}{30},
  \bibinfo{pages}{347--78; quiz 379--80}.
\newblock \URLprefix \url{https://www.ncbi.nlm.nih.gov/pubmed/20601558},
  \DOIprefix\doi{10.1177/1090820X10373357}.
\bibitem[{Beule(2010)}]{Beule2010}
\bibinfo{author}{Beule, A.G.}, \bibinfo{year}{2010}.
\newblock \bibinfo{title}{{Physiology and pathophysiology of respiratory mucosa
  of the nose and the paranasal sinuses.}}
\newblock \bibinfo{journal}{GMS current topics in otorhinolaryngology, head and
  neck surgery} \DOIprefix\doi{10.3205/cto000071}.
\bibitem[{Burgos et~al.(2017)Burgos, Sanmiguel-Rojas, del Pino,
  Sevilla-Garc{\'{i}}a and Esteban-Ortega}]{Burgos2017}
\bibinfo{author}{Burgos, M.A.}, \bibinfo{author}{Sanmiguel-Rojas, E.},
  \bibinfo{author}{del Pino, C.}, \bibinfo{author}{Sevilla-Garc{\'{i}}a, M.A.},
  \bibinfo{author}{Esteban-Ortega, F.}, \bibinfo{year}{2017}.
\newblock \bibinfo{title}{{New CFD tools to evaluate nasal airflow}}.
\newblock \bibinfo{journal}{European Archives of Oto-Rhino-Laryngology}
  \bibinfo{volume}{274}, \bibinfo{pages}{3121--3128}.
\newblock \DOIprefix\doi{10.1007/s00405-017-4611-y}.
\bibitem[{Chandra et~al.(2009)Chandra, Patadia and Raviv}]{Chandra2009}
\bibinfo{author}{Chandra, R.K.}, \bibinfo{author}{Patadia, M.O.},
  \bibinfo{author}{Raviv, J.}, \bibinfo{year}{2009}.
\newblock \bibinfo{title}{Diagnosis of nasal airway obstruction}.
\newblock \bibinfo{journal}{Otolaryngol Clin North Am} \bibinfo{volume}{42},
  \bibinfo{pages}{207--25, vii}.
\newblock \URLprefix \url{https://www.ncbi.nlm.nih.gov/pubmed/19328887},
  \DOIprefix\doi{10.1016/j.otc.2009.01.004}.
\bibitem[{Chen and Huang(2019)}]{Chen2019}
\bibinfo{author}{Chen, Y.Y.}, \bibinfo{author}{Huang, T.C.},
  \bibinfo{year}{2019}.
\newblock \bibinfo{title}{Outcome of septoplasty with inferior turbinectomy as
  an in-patient or out-patient procedure}.
\newblock \bibinfo{journal}{Sci Rep} \bibinfo{volume}{9},
  \bibinfo{pages}{7573}.
\newblock \URLprefix \url{https://www.ncbi.nlm.nih.gov/pubmed/31110327},
  \DOIprefix\doi{10.1038/s41598-019-44107-4}.
\bibitem[{Choi et~al.(2017)Choi, Cho, Choi, Zhang, Kim, Han, Kim, Kim, Rhee and
  Won}]{Choi2017}
\bibinfo{author}{Choi, K.Y.}, \bibinfo{author}{Cho, S.W.},
  \bibinfo{author}{Choi, J.J.}, \bibinfo{author}{Zhang, Y.L.},
  \bibinfo{author}{Kim, D.W.}, \bibinfo{author}{Han, D.H.},
  \bibinfo{author}{Kim, H.J.}, \bibinfo{author}{Kim, D.Y.},
  \bibinfo{author}{Rhee, C.S.}, \bibinfo{author}{Won, T.B.},
  \bibinfo{year}{2017}.
\newblock \bibinfo{title}{Healing of the nasal septal mucosa in an experimental
  rabbit model of mucosal injury}.
\newblock \bibinfo{journal}{World J Otorhinolaryngol Head Neck Surg}
  \bibinfo{volume}{3}, \bibinfo{pages}{17--23}.
\newblock \URLprefix \url{https://www.ncbi.nlm.nih.gov/pubmed/29204575},
  \DOIprefix\doi{10.1016/j.wjorl.2017.02.004}.
\bibitem[{Deniz et~al.(2014)Deniz, Ciftci, Isik, Demirel and
  Gultekin}]{Deniz2014}
\bibinfo{author}{Deniz, M.}, \bibinfo{author}{Ciftci, Z.},
  \bibinfo{author}{Isik, A.}, \bibinfo{author}{Demirel, O.B.},
  \bibinfo{author}{Gultekin, E.}, \bibinfo{year}{2014}.
\newblock \bibinfo{title}{The impact of different nasal packings on
  postoperative complications}.
\newblock \bibinfo{journal}{Am J Otolaryngol} \bibinfo{volume}{35},
  \bibinfo{pages}{554--7}.
\newblock \URLprefix \url{https://www.ncbi.nlm.nih.gov/pubmed/24943408},
  \DOIprefix\doi{10.1016/j.amjoto.2014.04.001}.
\bibitem[{Derin et~al.(2016)Derin, Sahan, Deveer, Erdogan, Tetiker and
  Koseoglu}]{Derin2016}
\bibinfo{author}{Derin, S.}, \bibinfo{author}{Sahan, M.},
  \bibinfo{author}{Deveer, M.}, \bibinfo{author}{Erdogan, S.},
  \bibinfo{author}{Tetiker, H.}, \bibinfo{author}{Koseoglu, S.},
  \bibinfo{year}{2016}.
\newblock \bibinfo{title}{The causes of persistent and recurrent nasal
  obstruction after primary septoplasty}.
\newblock \bibinfo{journal}{J Craniofac Surg} \bibinfo{volume}{27},
  \bibinfo{pages}{828--30}.
\newblock \URLprefix \url{https://www.ncbi.nlm.nih.gov/pubmed/27171946},
  \DOIprefix\doi{10.1097/SCS.0000000000002505}.
\bibitem[{Dong et~al.(2018)Dong, Shang, Inthavong, Chan and Tu}]{Dong2018}
\bibinfo{author}{Dong, J.}, \bibinfo{author}{Shang, Y.},
  \bibinfo{author}{Inthavong, K.}, \bibinfo{author}{Chan, H.K.},
  \bibinfo{author}{Tu, J.}, \bibinfo{year}{2018}.
\newblock \bibinfo{title}{{Partitioning of dispersed nanoparticles in a
  realistic nasal passage for targeted drug delivery}}.
\newblock \bibinfo{journal}{International Journal of Pharmaceutics}
  \DOIprefix\doi{10.1016/j.ijpharm.2018.03.046}.
\bibitem[{Doorly et~al.(2008)Doorly, Taylor and Schroter}]{Doorly2008}
\bibinfo{author}{Doorly, D.J.}, \bibinfo{author}{Taylor, D.J.},
  \bibinfo{author}{Schroter, R.C.}, \bibinfo{year}{2008}.
\newblock \bibinfo{title}{{Mechanics of airflow in the human nasal airways}}.
\newblock \bibinfo{journal}{Respiratory Physiology and Neurobiology}
  \bibinfo{volume}{163}, \bibinfo{pages}{100--110}.
\newblock \DOIprefix\doi{10.1016/j.resp.2008.07.027}.
\bibitem[{Elad et~al.(2008)Elad, Wolf and Keck}]{elad2008air}
\bibinfo{author}{Elad, D.}, \bibinfo{author}{Wolf, M.}, \bibinfo{author}{Keck,
  T.}, \bibinfo{year}{2008}.
\newblock \bibinfo{title}{Air-conditioning in the human nasal cavity}.
\newblock \bibinfo{journal}{Respiratory physiology \& neurobiology}
  \bibinfo{volume}{163}, \bibinfo{pages}{121--127}.
\bibitem[{Frank-Ito et~al.(2019)Frank-Ito, Kimbell, Borojeni, Garcia and
  Rhee}]{FrankIto2019}
\bibinfo{author}{Frank-Ito, D.O.}, \bibinfo{author}{Kimbell, J.S.},
  \bibinfo{author}{Borojeni, A.A.T.}, \bibinfo{author}{Garcia, G.J.M.},
  \bibinfo{author}{Rhee, J.S.}, \bibinfo{year}{2019}.
\newblock \bibinfo{title}{A hierarchical stepwise approach to evaluate nasal
  patency after virtual surgery for nasal airway obstruction}.
\newblock \bibinfo{journal}{Clin Biomech (Bristol, Avon)} \bibinfo{volume}{61},
  \bibinfo{pages}{172--180}.
\newblock \URLprefix \url{https://www.ncbi.nlm.nih.gov/pubmed/30594764},
  \DOIprefix\doi{10.1016/j.clinbiomech.2018.12.014}.
\bibitem[{Frank-Ito et~al.(2016)Frank-Ito, Wofford, Schroeter and
  Kimbell}]{frank2016influence}
\bibinfo{author}{Frank-Ito, D.O.}, \bibinfo{author}{Wofford, M.},
  \bibinfo{author}{Schroeter, J.D.}, \bibinfo{author}{Kimbell, J.S.},
  \bibinfo{year}{2016}.
\newblock \bibinfo{title}{Influence of mesh density on airflow and particle
  deposition in sinonasal airway modeling}.
\newblock \bibinfo{journal}{Journal of aerosol medicine and pulmonary drug
  delivery} \bibinfo{volume}{29}, \bibinfo{pages}{46--56}.
\bibitem[{Gaberino et~al.(2017)Gaberino, Rhee and Garcia}]{Gaberino2017}
\bibinfo{author}{Gaberino, C.}, \bibinfo{author}{Rhee, J.S.},
  \bibinfo{author}{Garcia, G.J.}, \bibinfo{year}{2017}.
\newblock \bibinfo{title}{{Estimates of nasal airflow at the nasal cycle
  mid-point improve the correlation between objective and subjective measures
  of nasal patency}}.
\newblock \bibinfo{journal}{Respiratory Physiology and Neurobiology}
  \DOIprefix\doi{10.1016/j.resp.2017.01.004}.
\bibitem[{Garcia et~al.(2015)Garcia, Schroeter and Kimbell}]{Garcia2015}
\bibinfo{author}{Garcia, G.J.}, \bibinfo{author}{Schroeter, J.D.},
  \bibinfo{author}{Kimbell, J.S.}, \bibinfo{year}{2015}.
\newblock \bibinfo{title}{{Olfactory deposition of inhaled nanoparticles in
  humans}}.
\newblock \bibinfo{journal}{Inhalation Toxicology}
  \DOIprefix\doi{10.3109/08958378.2015.1066904}.
\bibitem[{Ge et~al.(2012)Ge, Inthavong and Tu}]{Ge2012}
\bibinfo{author}{Ge, Q.J.}, \bibinfo{author}{Inthavong, K.},
  \bibinfo{author}{Tu, J.Y.}, \bibinfo{year}{2012}.
\newblock \bibinfo{title}{{Local deposition fractions of ultrafine particles in
  a human nasal-sinus cavity CFD model}}.
\newblock \bibinfo{journal}{Inhalation Toxicology}
  \DOIprefix\doi{10.3109/08958378.2012.694494}.
\bibitem[{Henriquez et~al.(2013)Henriquez, Schlosser, Mace, Smith and
  Soler}]{Henriquez2013}
\bibinfo{author}{Henriquez, O.A.}, \bibinfo{author}{Schlosser, R.J.},
  \bibinfo{author}{Mace, J.C.}, \bibinfo{author}{Smith, T.L.},
  \bibinfo{author}{Soler, Z.M.}, \bibinfo{year}{2013}.
\newblock \bibinfo{title}{Impact of synechiae after endoscopic sinus surgery on
  long-term outcomes in chronic rhinosinusitis}.
\newblock \bibinfo{journal}{Laryngoscope} \bibinfo{volume}{123},
  \bibinfo{pages}{2615--9}.
\newblock \URLprefix \url{https://www.ncbi.nlm.nih.gov/pubmed/23670876},
  \DOIprefix\doi{10.1002/lary.24150}.
\bibitem[{Inthavong(2020)}]{Inthavong2020a}
\bibinfo{author}{Inthavong, K.}, \bibinfo{year}{2020}.
\newblock \bibinfo{title}{{From indoor exposure to inhaled particle deposition:
  A multiphase journey of inhaled particles}}.
\newblock \bibinfo{journal}{Experimental and Computational Multiphase Flow}
  \DOIprefix\doi{10.1007/s42757-019-0046-6}.
\bibitem[{Inthavong et~al.(2018)Inthavong, Chetty, Shang and
  Tu}]{Inthavong2018}
\bibinfo{author}{Inthavong, K.}, \bibinfo{author}{Chetty, A.},
  \bibinfo{author}{Shang, Y.}, \bibinfo{author}{Tu, J.}, \bibinfo{year}{2018}.
\newblock \bibinfo{title}{{Examining mesh independence for flow dynamics in the
  human nasal cavity}}.
\newblock \bibinfo{journal}{Computers in Biology and Medicine}
  \DOIprefix\doi{10.1016/j.compbiomed.2018.09.010}.
\bibitem[{Inthavong et~al.(2019)Inthavong, Ma, Shang, Dong, Chetty, Tu and
  Frank-Ito}]{Inthavong2019}
\bibinfo{author}{Inthavong, K.}, \bibinfo{author}{Ma, J.},
  \bibinfo{author}{Shang, Y.}, \bibinfo{author}{Dong, J.},
  \bibinfo{author}{Chetty, A.S.}, \bibinfo{author}{Tu, J.},
  \bibinfo{author}{Frank-Ito, D.}, \bibinfo{year}{2019}.
\newblock \bibinfo{title}{{Geometry and airflow dynamics analysis in the nasal
  cavity during inhalation}}.
\newblock \bibinfo{journal}{Clinical Biomechanics}
  \DOIprefix\doi{10.1016/j.clinbiomech.2017.10.006}.
\bibitem[{Ito et~al.(2017)Ito, Mitsumune, Kuga, Phuong, Tani and
  Inthavong}]{Ito2017}
\bibinfo{author}{Ito, K.}, \bibinfo{author}{Mitsumune, K.},
  \bibinfo{author}{Kuga, K.}, \bibinfo{author}{Phuong, N.L.},
  \bibinfo{author}{Tani, K.}, \bibinfo{author}{Inthavong, K.},
  \bibinfo{year}{2017}.
\newblock \bibinfo{title}{Prediction of convective heat transfer coefficients
  for the upper respiratory tracts of rat, dog, monkey, and humans}.
\newblock \bibinfo{journal}{Indoor and Built Environment} \bibinfo{volume}{26},
  \bibinfo{pages}{828--840}.
\bibitem[{Jo et~al.(2015)Jo, Chung and Na}]{Jo2015}
\bibinfo{author}{Jo, G.}, \bibinfo{author}{Chung, S.K.}, \bibinfo{author}{Na,
  Y.}, \bibinfo{year}{2015}.
\newblock \bibinfo{title}{{Numerical study of the effect of the nasal cycle on
  unilateral nasal resistance}}.
\newblock \bibinfo{journal}{Respiratory Physiology and Neurobiology}
  \DOIprefix\doi{10.1016/j.resp.2015.08.006}.
\bibitem[{Joe~Jacob et~al.(2011)Joe~Jacob, George, Preethi and
  Arunraj}]{JoeJacob2011}
\bibinfo{author}{Joe~Jacob, K.}, \bibinfo{author}{George, S.},
  \bibinfo{author}{Preethi, S.}, \bibinfo{author}{Arunraj, V.S.},
  \bibinfo{year}{2011}.
\newblock \bibinfo{title}{A comparative study between endoscopic middle meatal
  antrostomy and caldwell-luc surgery in the treatment of chronic maxillary
  sinusitis}.
\newblock \bibinfo{journal}{Indian journal of otolaryngology and head and neck
  surgery : official publication of the Association of Otolaryngologists of
  India} \bibinfo{volume}{63}, \bibinfo{pages}{214--219}.
\newblock \URLprefix \url{https://www.ncbi.nlm.nih.gov/pubmed/22754797
  https://www.ncbi.nlm.nih.gov/pmc/articles/PMC3138946/},
  \DOIprefix\doi{10.1007/s12070-011-0262-2}.
\bibitem[{Jones(1999)}]{Jones1999}
\bibinfo{author}{Jones, N.S.}, \bibinfo{year}{1999}.
\newblock \bibinfo{title}{Nasal manifestations of rheumatic diseases}.
\newblock \bibinfo{journal}{Ann Rheum Dis} \bibinfo{volume}{58},
  \bibinfo{pages}{589--90}.
\newblock \URLprefix \url{https://www.ncbi.nlm.nih.gov/pubmed/10491354},
  \DOIprefix\doi{10.1136/ard.58.10.589}.
\bibitem[{Joshi et~al.(2019)Joshi, Riley and Kacker}]{Joshi2019}
\bibinfo{author}{Joshi, R.R.}, \bibinfo{author}{Riley, C.A.},
  \bibinfo{author}{Kacker, A.}, \bibinfo{year}{2019}.
\newblock \bibinfo{title}{Complication rates following septoplasty with
  inferior turbinate reduction}.
\newblock \bibinfo{journal}{Ochsner J} \bibinfo{volume}{19},
  \bibinfo{pages}{353--356}.
\newblock \URLprefix \url{https://www.ncbi.nlm.nih.gov/pubmed/31903059},
  \DOIprefix\doi{10.31486/toj.19.0002}.
\bibitem[{Keck et~al.(2000)Keck, Leiacker, Heinrich, K{\"u}hnemann and
  Rettinger}]{keck2000}
\bibinfo{author}{Keck, T.}, \bibinfo{author}{Leiacker, R.},
  \bibinfo{author}{Heinrich, A.}, \bibinfo{author}{K{\"u}hnemann, S.},
  \bibinfo{author}{Rettinger, G.}, \bibinfo{year}{2000}.
\newblock \bibinfo{title}{Humidity and temperature profile in the nasal
  cavity.}
\newblock \bibinfo{journal}{Rhinology} \bibinfo{volume}{38},
  \bibinfo{pages}{167--171}.
\bibitem[{Keh et~al.(2011)Keh, Facer, Yehia, Sandhu, Saleh and Anand}]{Keh2011}
\bibinfo{author}{Keh, S.M.}, \bibinfo{author}{Facer, P.},
  \bibinfo{author}{Yehia, A.}, \bibinfo{author}{Sandhu, G.},
  \bibinfo{author}{Saleh, H.A.}, \bibinfo{author}{Anand, P.},
  \bibinfo{year}{2011}.
\newblock \bibinfo{title}{The menthol and cold sensation receptor trpm8 in
  normal human nasal mucosa and rhinitis}.
\newblock \bibinfo{journal}{Rhinology} \bibinfo{volume}{49},
  \bibinfo{pages}{453--7}.
\newblock \URLprefix \url{https://www.ncbi.nlm.nih.gov/pubmed/21991571},
  \DOIprefix\doi{10.4193/Rhin11.089}.
\bibitem[{Kim et~al.(2014)Kim, Heo, Seo, Na and Chung}]{Kim2014}
\bibinfo{author}{Kim, S.K.}, \bibinfo{author}{Heo, G.E.}, \bibinfo{author}{Seo,
  A.}, \bibinfo{author}{Na, Y.}, \bibinfo{author}{Chung, S.K.},
  \bibinfo{year}{2014}.
\newblock \bibinfo{title}{{Correlation between nasal airflow characteristics
  and clinical relevance of nasal septal deviation to nasal airway
  obstruction}}.
\newblock \bibinfo{journal}{Respiratory Physiology and Neurobiology}
  \DOIprefix\doi{10.1016/j.resp.2013.12.010}.
\bibitem[{Lindemann et~al.(2013)Lindemann, Rettinger, Kroger and
  Sommer}]{Lindemann2013}
\bibinfo{author}{Lindemann, J.}, \bibinfo{author}{Rettinger, G.},
  \bibinfo{author}{Kroger, R.}, \bibinfo{author}{Sommer, F.},
  \bibinfo{year}{2013}.
\newblock \bibinfo{title}{Numerical simulation of airflow patterns in nose
  models with differently localized septal perforations}.
\newblock \bibinfo{journal}{Laryngoscope} \bibinfo{volume}{123},
  \bibinfo{pages}{2085--9}.
\newblock \URLprefix \url{https://www.ncbi.nlm.nih.gov/pubmed/23821431},
  \DOIprefix\doi{10.1002/lary.23653}.
\bibitem[{Liu et~al.(2015)Liu, Lu, Cheng, Chu, Lee, Wu and Wang}]{Liu2015}
\bibinfo{author}{Liu, S.C.}, \bibinfo{author}{Lu, H.H.},
  \bibinfo{author}{Cheng, L.H.}, \bibinfo{author}{Chu, Y.H.},
  \bibinfo{author}{Lee, F.P.}, \bibinfo{author}{Wu, C.C.},
  \bibinfo{author}{Wang, H.W.}, \bibinfo{year}{2015}.
\newblock \bibinfo{title}{Identification of the cold receptor trpm8 in the
  nasal mucosa}.
\newblock \bibinfo{journal}{Am J Rhinol Allergy} \bibinfo{volume}{29},
  \bibinfo{pages}{e112--6}.
\newblock \URLprefix \url{https://www.ncbi.nlm.nih.gov/pubmed/26163239},
  \DOIprefix\doi{10.2500/ajra.2015.29.4202}.
\bibitem[{Lu~Phuong et~al.(2018)Lu~Phuong, Dang~Khoa, Inthavong and
  Ito}]{lu2018part}
\bibinfo{author}{Lu~Phuong, N.}, \bibinfo{author}{Dang~Khoa, N.},
  \bibinfo{author}{Inthavong, K.}, \bibinfo{author}{Ito, K.},
  \bibinfo{year}{2018}.
\newblock \bibinfo{title}{Particle and inhalation exposure in human and monkey
  computational airway models}.
\newblock \bibinfo{journal}{Inhalation toxicology} \bibinfo{volume}{30},
  \bibinfo{pages}{416--428}.
\bibitem[{Na et~al.(2020)Na, Chung and Byun}]{Na2020}
\bibinfo{author}{Na, Y.}, \bibinfo{author}{Chung, S.K.}, \bibinfo{author}{Byun,
  S.}, \bibinfo{year}{2020}.
\newblock \bibinfo{title}{{Numerical study on the heat-recovery capacity of the
  human nasal cavity during expiration}}.
\newblock \bibinfo{journal}{Computers in Biology and Medicine}
  \bibinfo{volume}{126}, \bibinfo{pages}{103992}.
\newblock \URLprefix
  \url{http://www.sciencedirect.com/science/article/pii/S0010482520303231},
  \DOIprefix\doi{https://doi.org/10.1016/j.compbiomed.2020.103992}.
\bibitem[{Naghibzadeh et~al.(2011)Naghibzadeh, Peyvandi and
  Naghibzadeh}]{Naghibzadeh2011}
\bibinfo{author}{Naghibzadeh, B.}, \bibinfo{author}{Peyvandi, A.A.},
  \bibinfo{author}{Naghibzadeh, G.}, \bibinfo{year}{2011}.
\newblock \bibinfo{title}{Does post septoplasty nasal packing reduce
  complications?}
\newblock \bibinfo{journal}{Acta Med Iran} \bibinfo{volume}{49},
  \bibinfo{pages}{9--12}.
\newblock \URLprefix \url{https://www.ncbi.nlm.nih.gov/pubmed/21425063}.
\bibitem[{Nayak et~al.(1998)Nayak, Balakrishnan and Hazarika}]{Nayak1998}
\bibinfo{author}{Nayak, D.R.}, \bibinfo{author}{Balakrishnan, R.},
  \bibinfo{author}{Hazarika, P.}, \bibinfo{year}{1998}.
\newblock \bibinfo{title}{Prevention and management of synechia in pediatric
  endoscopic sinus surgery using dental wax plates}.
\newblock \bibinfo{journal}{Int J Pediatr Otorhinolaryngol}
  \bibinfo{volume}{46}, \bibinfo{pages}{171--8}.
\newblock \URLprefix \url{https://www.ncbi.nlm.nih.gov/pubmed/10190587},
  \DOIprefix\doi{10.1016/s0165-5876(98)00124-4}.
\bibitem[{Nishijima et~al.(2018)Nishijima, Kondo, Yamamoto, Nomura, Kikuta,
  Shimizu, Mizushima and Yamasoba}]{Nishijima2018}
\bibinfo{author}{Nishijima, H.}, \bibinfo{author}{Kondo, K.},
  \bibinfo{author}{Yamamoto, T.}, \bibinfo{author}{Nomura, T.},
  \bibinfo{author}{Kikuta, S.}, \bibinfo{author}{Shimizu, Y.},
  \bibinfo{author}{Mizushima, Y.}, \bibinfo{author}{Yamasoba, T.},
  \bibinfo{year}{2018}.
\newblock \bibinfo{title}{Influence of the location of nasal polyps on
  olfactory airflow and olfaction}.
\newblock \bibinfo{journal}{Int Forum Allergy Rhinol} \bibinfo{volume}{8},
  \bibinfo{pages}{695--706}.
\newblock \URLprefix \url{https://www.ncbi.nlm.nih.gov/pubmed/29394000},
  \DOIprefix\doi{10.1002/alr.22089}.
\bibitem[{Rettinger and Kirsche(2006)}]{Rettinger2006}
\bibinfo{author}{Rettinger, G.}, \bibinfo{author}{Kirsche, H.},
  \bibinfo{year}{2006}.
\newblock \bibinfo{title}{Complications in septoplasty}.
\newblock \bibinfo{journal}{Facial Plast Surg} \bibinfo{volume}{22},
  \bibinfo{pages}{289--97}.
\newblock \URLprefix \url{https://www.ncbi.nlm.nih.gov/pubmed/17131271},
  \DOIprefix\doi{10.1055/s-2006-954847}.
\bibitem[{Salati et~al.(2021)Salati, Bartley, Yazdi, Jermy and
  White}]{Salati2021}
\bibinfo{author}{Salati, H.}, \bibinfo{author}{Bartley, J.},
  \bibinfo{author}{Yazdi, S.G.}, \bibinfo{author}{Jermy, M.},
  \bibinfo{author}{White, D.E.}, \bibinfo{year}{2021}.
\newblock \bibinfo{title}{{Neti pot irrigation volume filling simulation using
  anatomically accurate in-vivo nasal airway geometry}}.
\newblock \bibinfo{journal}{Respiratory Physiology and Neurobiology}
  \DOIprefix\doi{10.1016/j.resp.2020.103580}.
\bibitem[{Shang and Inthavong(2019)}]{shang2019}
\bibinfo{author}{Shang, Y.}, \bibinfo{author}{Inthavong, K.},
  \bibinfo{year}{2019}.
\newblock \bibinfo{title}{Numerical assessment of ambient inhaled micron
  particle deposition in a human nasal cavity}.
\newblock \bibinfo{journal}{Experimental and Computational Multiphase Flow}
  \bibinfo{volume}{1}, \bibinfo{pages}{109--115}.
\bibitem[{Singha and Sinhamahapatra(2010)}]{SINGHA2010757}
\bibinfo{author}{Singha, S.}, \bibinfo{author}{Sinhamahapatra, K.P.},
  \bibinfo{year}{2010}.
\newblock \bibinfo{title}{{Flow past a circular cylinder between parallel walls
  at low Reynolds numbers}}.
\newblock \bibinfo{journal}{Ocean Engineering} \bibinfo{volume}{37},
  \bibinfo{pages}{757--769}.
\newblock \URLprefix
  \url{https://www.sciencedirect.com/science/article/pii/S0029801810000570},
  \DOIprefix\doi{https://doi.org/10.1016/j.oceaneng.2010.02.012}.
\bibitem[{Stewart et~al.(2004)Stewart, Witsell, Smith, Weaver, Yueh and
  Hannley}]{Stewart2004}
\bibinfo{author}{Stewart, M.G.}, \bibinfo{author}{Witsell, D.L.},
  \bibinfo{author}{Smith, T.L.}, \bibinfo{author}{Weaver, E.M.},
  \bibinfo{author}{Yueh, B.}, \bibinfo{author}{Hannley, M.T.},
  \bibinfo{year}{2004}.
\newblock \bibinfo{title}{Development and validation of the nasal obstruction
  symptom evaluation (nose) scale}.
\newblock \bibinfo{journal}{Otolaryngol Head Neck Surg} \bibinfo{volume}{130},
  \bibinfo{pages}{157--63}.
\newblock \URLprefix \url{https://www.ncbi.nlm.nih.gov/pubmed/14990910},
  \DOIprefix\doi{10.1016/j.otohns.2003.09.016}.
\bibitem[{Sullivan et~al.(2014)Sullivan, Garcia, Frank-Ito, Kimbell and
  Rhee}]{Sullivan2014}
\bibinfo{author}{Sullivan, C.D.}, \bibinfo{author}{Garcia, G.J.},
  \bibinfo{author}{Frank-Ito, D.O.}, \bibinfo{author}{Kimbell, J.S.},
  \bibinfo{author}{Rhee, J.S.}, \bibinfo{year}{2014}.
\newblock \bibinfo{title}{Perception of better nasal patency correlates with
  increased mucosal cooling after surgery for nasal obstruction}.
\newblock \bibinfo{journal}{Otolaryngol Head Neck Surg} \bibinfo{volume}{150},
  \bibinfo{pages}{139--47}.
\newblock \URLprefix \url{https://www.ncbi.nlm.nih.gov/pubmed/24154749},
  \DOIprefix\doi{10.1177/0194599813509776}.
\bibitem[{Tian et~al.(2017)Tian, Shang, Dong, Inthavong and Tu}]{Tian2017}
\bibinfo{author}{Tian, L.}, \bibinfo{author}{Shang, Y.}, \bibinfo{author}{Dong,
  J.}, \bibinfo{author}{Inthavong, K.}, \bibinfo{author}{Tu, J.},
  \bibinfo{year}{2017}.
\newblock \bibinfo{title}{{Human nasal olfactory deposition of inhaled
  nanoparticles at low to moderate breathing rate}}.
\newblock \bibinfo{journal}{Journal of Aerosol Science}
  \DOIprefix\doi{10.1016/j.jaerosci.2017.08.006}.
\bibitem[{Van~Strien et~al.(2021)Van~Strien, Shrestha, Gabriel, Lappas,
  Fletcher, Singh and Inthavong}]{van2021pressure}
\bibinfo{author}{Van~Strien, J.}, \bibinfo{author}{Shrestha, K.},
  \bibinfo{author}{Gabriel, S.}, \bibinfo{author}{Lappas, P.},
  \bibinfo{author}{Fletcher, D.F.}, \bibinfo{author}{Singh, N.},
  \bibinfo{author}{Inthavong, K.}, \bibinfo{year}{2021}.
\newblock \bibinfo{title}{Pressure distribution and flow dynamics in a nasal
  airway using a scale resolving simulation}.
\newblock \bibinfo{journal}{Physics of Fluids} \bibinfo{volume}{33},
  \bibinfo{pages}{011907}.
\bibitem[{Vartiainen and Nuutinen(1992)}]{Vartiainen1992}
\bibinfo{author}{Vartiainen, E.}, \bibinfo{author}{Nuutinen, J.},
  \bibinfo{year}{1992}.
\newblock \bibinfo{title}{Head and neck manifestations of wegener's
  granulomatosis}.
\newblock \bibinfo{journal}{Ear Nose Throat J} \bibinfo{volume}{71},
  \bibinfo{pages}{423--4; 427--8}.
\newblock \URLprefix \url{https://www.ncbi.nlm.nih.gov/pubmed/1425382}.
\bibitem[{Wang et~al.(2016)Wang, Chen, Wang, Chen and Deng}]{Wang2016}
\bibinfo{author}{Wang, T.}, \bibinfo{author}{Chen, D.}, \bibinfo{author}{Wang,
  P.H.}, \bibinfo{author}{Chen, J.}, \bibinfo{author}{Deng, J.},
  \bibinfo{year}{2016}.
\newblock \bibinfo{title}{Investigation on the nasal airflow characteristics of
  anterior nasal cavity stenosis}.
\newblock \bibinfo{journal}{Braz J Med Biol Res} \bibinfo{volume}{49},
  \bibinfo{pages}{e5182}.
\newblock \URLprefix \url{https://www.ncbi.nlm.nih.gov/pubmed/27533764},
  \DOIprefix\doi{10.1590/1414-431X20165182}.
\bibitem[{Xiong et~al.(2008)Xiong, Zhan, Jiang, Li, Rong and Xu}]{Xiong2008}
\bibinfo{author}{Xiong, G.X.}, \bibinfo{author}{Zhan, J.M.},
  \bibinfo{author}{Jiang, H.Y.}, \bibinfo{author}{Li, J.F.},
  \bibinfo{author}{Rong, L.W.}, \bibinfo{author}{Xu, G.}, \bibinfo{year}{2008}.
\newblock \bibinfo{title}{Computational fluid dynamics simulation of airflow in
  the normal nasal cavity and paranasal sinuses}.
\newblock \bibinfo{journal}{Am J Rhinol} \bibinfo{volume}{22},
  \bibinfo{pages}{477--82}.
\newblock \URLprefix \url{https://www.ncbi.nlm.nih.gov/pubmed/18954506},
  \DOIprefix\doi{10.2500/ajr.2008.22.3211}.
\bibitem[{Young et~al.(1997)Young, Finn and Kim}]{Young1997}
\bibinfo{author}{Young, T.}, \bibinfo{author}{Finn, L.}, \bibinfo{author}{Kim,
  H.}, \bibinfo{year}{1997}.
\newblock \bibinfo{title}{Nasal obstruction as a risk factor for
  sleep-disordered breathing. the university of wisconsin sleep and respiratory
  research group}.
\newblock \bibinfo{journal}{J Allergy Clin Immunol} \bibinfo{volume}{99},
  \bibinfo{pages}{S757--62}.
\newblock \URLprefix \url{https://www.ncbi.nlm.nih.gov/pubmed/9042068},
  \DOIprefix\doi{10.1016/s0091-6749(97)70124-6}.
\bibitem[{Zhao et~al.(2011)Zhao, Blacker, Luo, Bryant and Jiang}]{Zhao2011}
\bibinfo{author}{Zhao, K.}, \bibinfo{author}{Blacker, K.},
  \bibinfo{author}{Luo, Y.}, \bibinfo{author}{Bryant, B.},
  \bibinfo{author}{Jiang, J.}, \bibinfo{year}{2011}.
\newblock \bibinfo{title}{Perceiving nasal patency through mucosal cooling
  rather than air temperature or nasal resistance}.
\newblock \bibinfo{journal}{PLoS One} \bibinfo{volume}{6},
  \bibinfo{pages}{e24618}.
\newblock \URLprefix \url{https://www.ncbi.nlm.nih.gov/pubmed/22022361},
  \DOIprefix\doi{10.1371/journal.pone.0024618}.
\bibitem[{Zhao et~al.(2014)Zhao, Jiang, Blacker, Lyman, Dalton, Cowart and
  Pribitkin}]{Zhao2014}
\bibinfo{author}{Zhao, K.}, \bibinfo{author}{Jiang, J.},
  \bibinfo{author}{Blacker, K.}, \bibinfo{author}{Lyman, B.},
  \bibinfo{author}{Dalton, P.}, \bibinfo{author}{Cowart, B.J.},
  \bibinfo{author}{Pribitkin, E.A.}, \bibinfo{year}{2014}.
\newblock \bibinfo{title}{Regional peak mucosal cooling predicts the perception
  of nasal patency}.
\newblock \bibinfo{journal}{Laryngoscope} \bibinfo{volume}{124},
  \bibinfo{pages}{589--95}.
\newblock \URLprefix \url{https://www.ncbi.nlm.nih.gov/pubmed/23775640},
  \DOIprefix\doi{10.1002/lary.24265}.
\bibitem[{Zhao et~al.(2004)Zhao, Scherer, Hajiloo and Dalton}]{Zhao2004}
\bibinfo{author}{Zhao, K.}, \bibinfo{author}{Scherer, P.W.},
  \bibinfo{author}{Hajiloo, S.A.}, \bibinfo{author}{Dalton, P.},
  \bibinfo{year}{2004}.
\newblock \bibinfo{title}{Effect of anatomy on human nasal air flow and odorant
  transport patterns: implications for olfaction}.
\newblock \bibinfo{journal}{Chem Senses} \bibinfo{volume}{29},
  \bibinfo{pages}{365--79}.
\newblock \URLprefix \url{https://www.ncbi.nlm.nih.gov/pubmed/15201204},
  \DOIprefix\doi{10.1093/chemse/bjh033}.
\bibitem[{Zhu et~al.(2012)Zhu, Lee, Lim, Gordon and Wang}]{Zhu2012}
\bibinfo{author}{Zhu, J.H.}, \bibinfo{author}{Lee, H.P.}, \bibinfo{author}{Lim,
  K.M.}, \bibinfo{author}{Gordon, B.R.}, \bibinfo{author}{Wang, D.Y.},
  \bibinfo{year}{2012}.
\newblock \bibinfo{title}{{Effect of accessory ostia on maxillary sinus
  ventilation: A computational fluid dynamics (CFD) study}}.
\newblock \bibinfo{journal}{Respiratory Physiology and Neurobiology}
  \DOIprefix\doi{10.1016/j.resp.2012.06.026}.
\bibitem[{Zhu et~al.(2011)Zhu, Lee, Lim, Lee et~al.}]{zhu2011eval}
\bibinfo{author}{Zhu, J.H.}, \bibinfo{author}{Lee, H.P.}, \bibinfo{author}{Lim,
  K.M.}, \bibinfo{author}{Lee, S.J.}, et~al., \bibinfo{year}{2011}.
\newblock \bibinfo{title}{Evaluation and comparison of nasal airway flow
  patterns among three subjects from caucasian, chinese and indian ethnic
  groups using computational fluid dynamics simulation}.
\newblock \bibinfo{journal}{Respiratory physiology \& neurobiology}
  \bibinfo{volume}{175}, \bibinfo{pages}{62--69}.

\end{thebibliography}

\normalsize
\newgeometry{margin=1.5cm}
\clearpage
\begin{figure}
	\centering
	\includegraphics[width=0.75\linewidth]{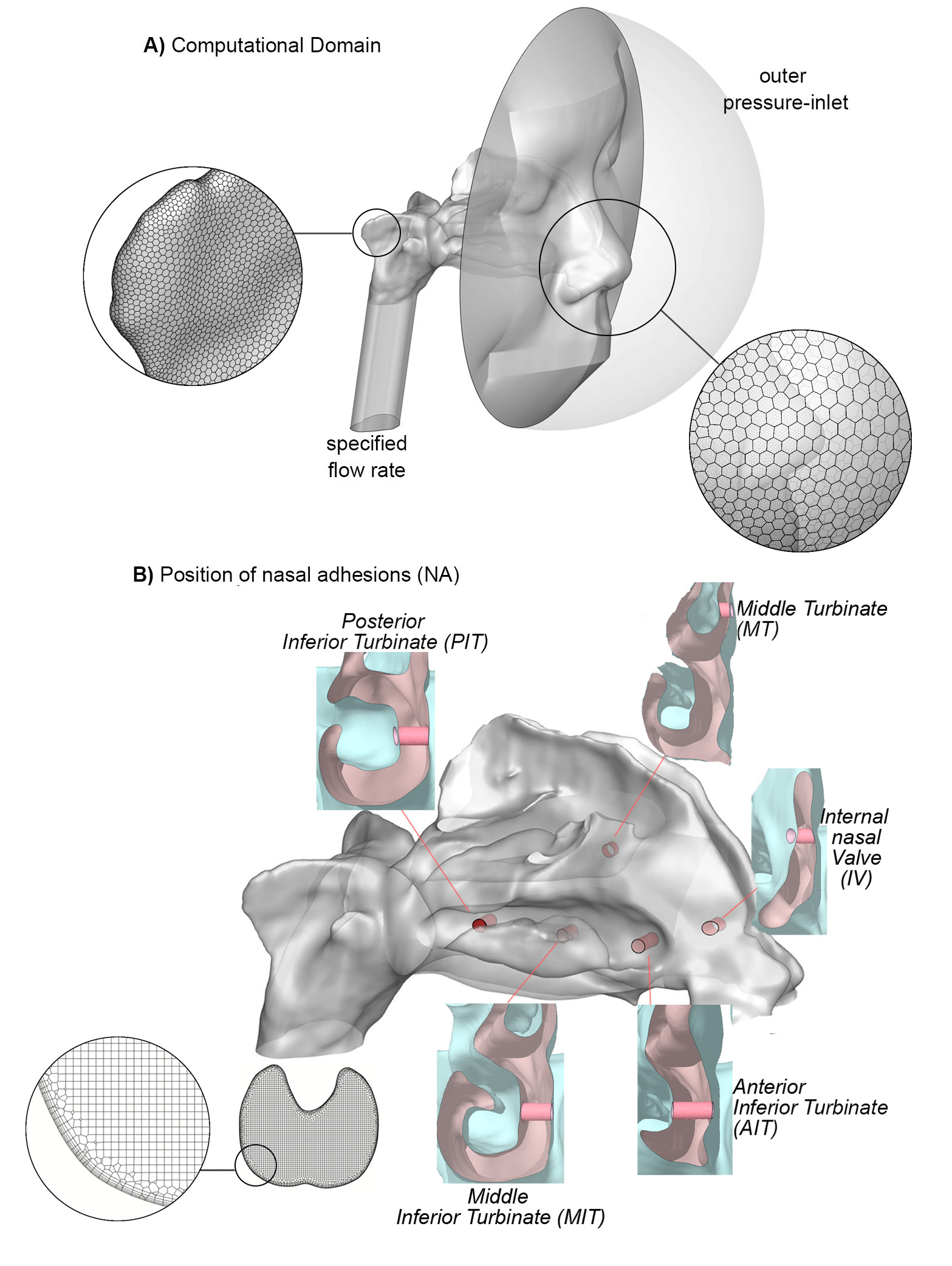}
	\caption{A) Computational domain depicting the outer face and nose. B) Right lateral-oblique view of the nasal cavity showing the locations of virtual nasal adhesions within the right cavity, combined for all individual models and displayed transparently in red. Three models were created with an adhesion between the inferior turbinate and septum: anterior (at the interior turbinate head, labelled AIT), middle (mid-inferior turbinate, labelled MIT) and posterior inferior turbinate (at the posterior end of the inferior turbinate, labelled PIT); one model had  an adhesion between the middle turbinate (MT) head and septum; and the final model located the adhesion high within the internal nasal valve (IV), between the upper lateral cartilage and septum.
}
	\label{fig:location}
\end{figure}

\clearpage
\begin{figure}
	\centering
	\includegraphics[width=1\linewidth]{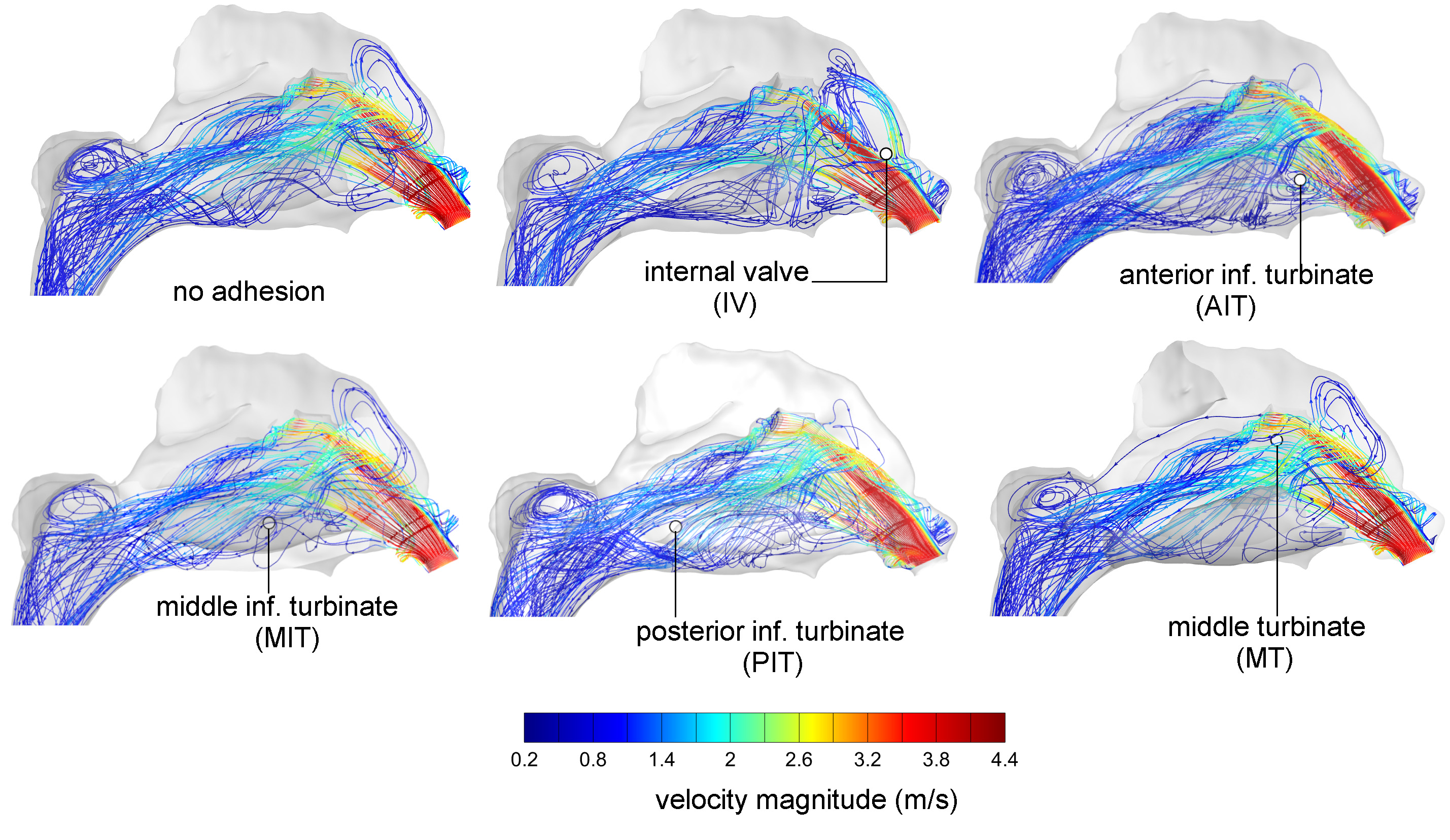}
	\caption{Streamline analysis depicting nasal airflow in normal and nasal adhesion (NA) models, shown from the right lateral view in the right nasal cavity. NA are shown as a black-outlined circle. IV, interval valve; AIT, anterior inferior turbinate; MIT, middle inferior turbinate; PIT, posterior inferior turbinate; MT, middle turbinate.}
	\label{fig:streamlines}
\end{figure}


\clearpage
\begin{figure}
	\centering
	\includegraphics[width=0.9\linewidth]{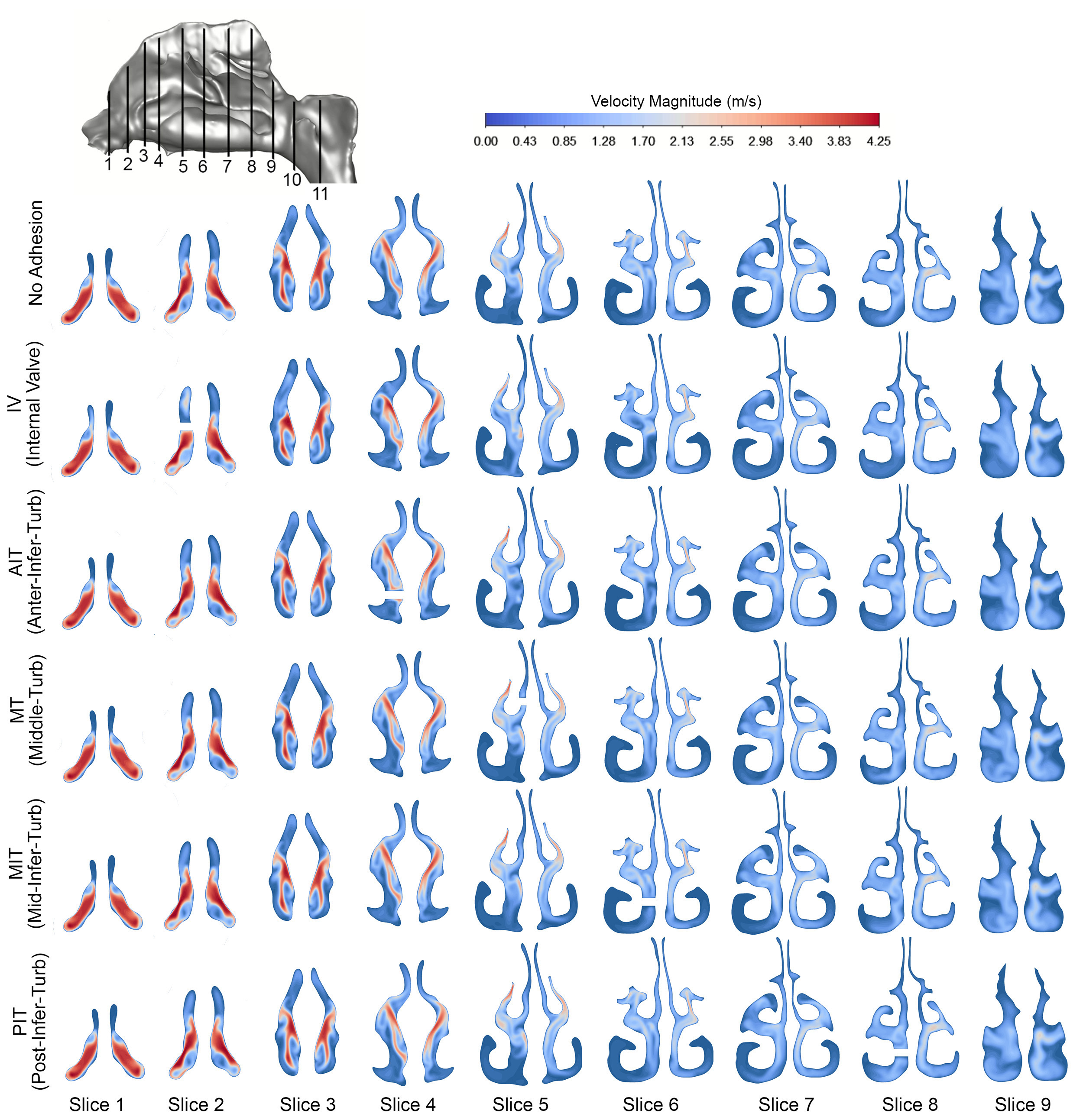}
	\caption{Velocity contours on the first nine coronal slices for all models. The adhesions correspond with slice locations where: internal valve (IV) adhesion is located at Slice 2; anterior-inferior turbinate (AIT) adhesion is located at Slice 4; middle turbinate (MT) adhesion is located at slice 5; middle-inferior turbinate (MIT) adhesion is located at slice 6; and posterior-inferior turbinate is located at slice 8.}
	\label{fig:vel-conts}
\end{figure}

\clearpage
\begin{figure}
	\centering
	\includegraphics[width=0.9\linewidth]{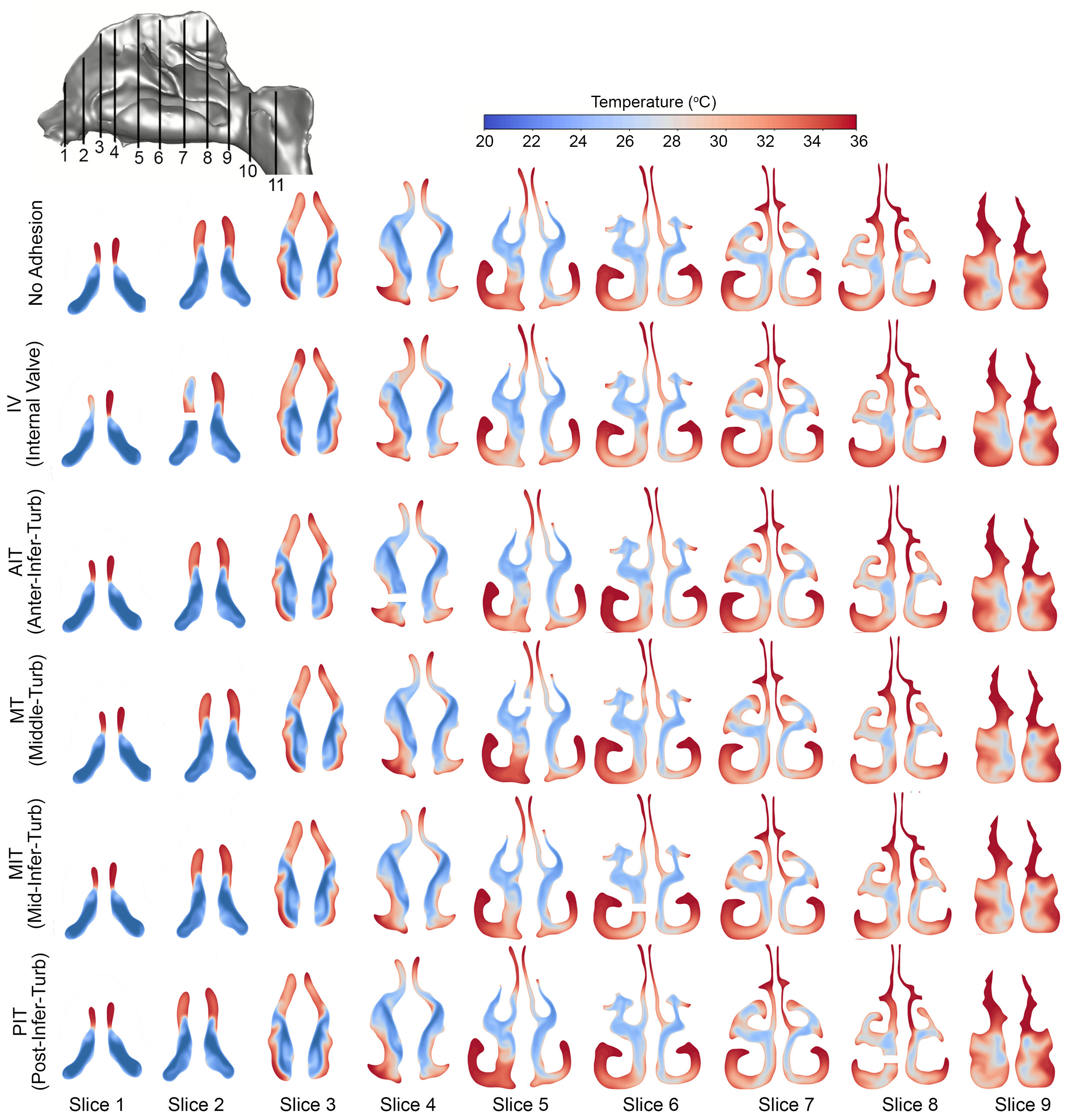}
	\caption{Temperature contours on the first nine coronal slices for all models. The adhesions correspond with slice locations where: internal valve (IV) adhesion is located at Slice 2; anterior-inferior turbinate (AIT) adhesion is located at Slice 4; middle turbinate (MT) adhesion is located at slice 5; middle-inferior turbinate (MIT) adhesion is located at slice 6; and posterior-inferior turbinate is located at slice 8.}
	\label{fig:temp-conts}
\end{figure}
\clearpage
\begin{figure}
	\centering
	\includegraphics[width=0.9\linewidth]{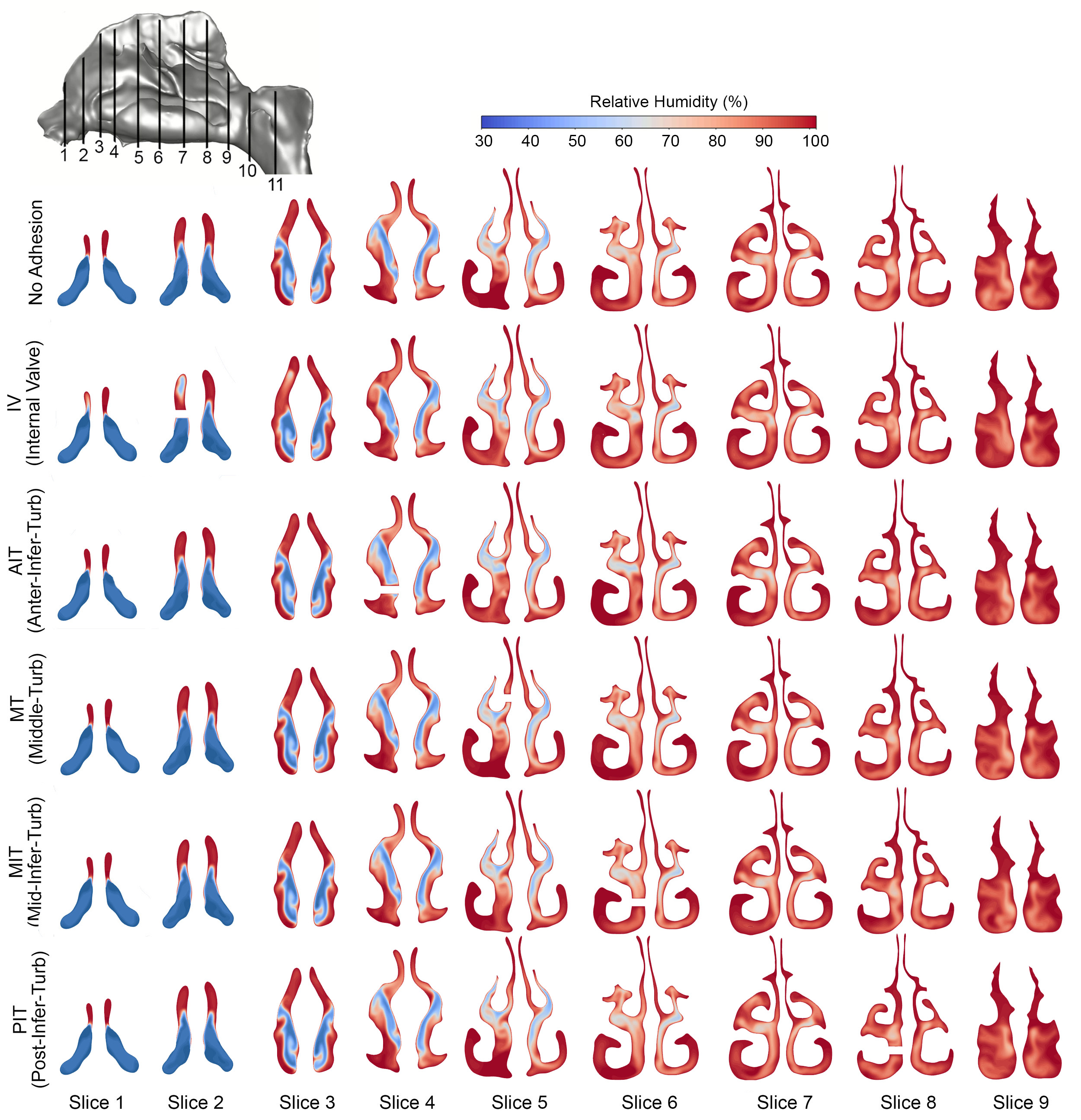}
	\caption{Relative humidity contours on the first nine coronal slices for all models. The adhesions correspond with slice locations where: internal valve (IV) adhesion is located at Slice 2; anterior-inferior turbinate (AIT) adhesion is located at Slice 4; middle turbinate (MT) adhesion is located at slice 5; middle-inferior turbinate (MIT) adhesion is located at slice 6; and posterior-inferior turbinate is located at slice 8.}
	\label{fig:rh-conts}
\end{figure}

\clearpage
\begin{figure}
	\centering
	\includegraphics[width=0.75\linewidth]{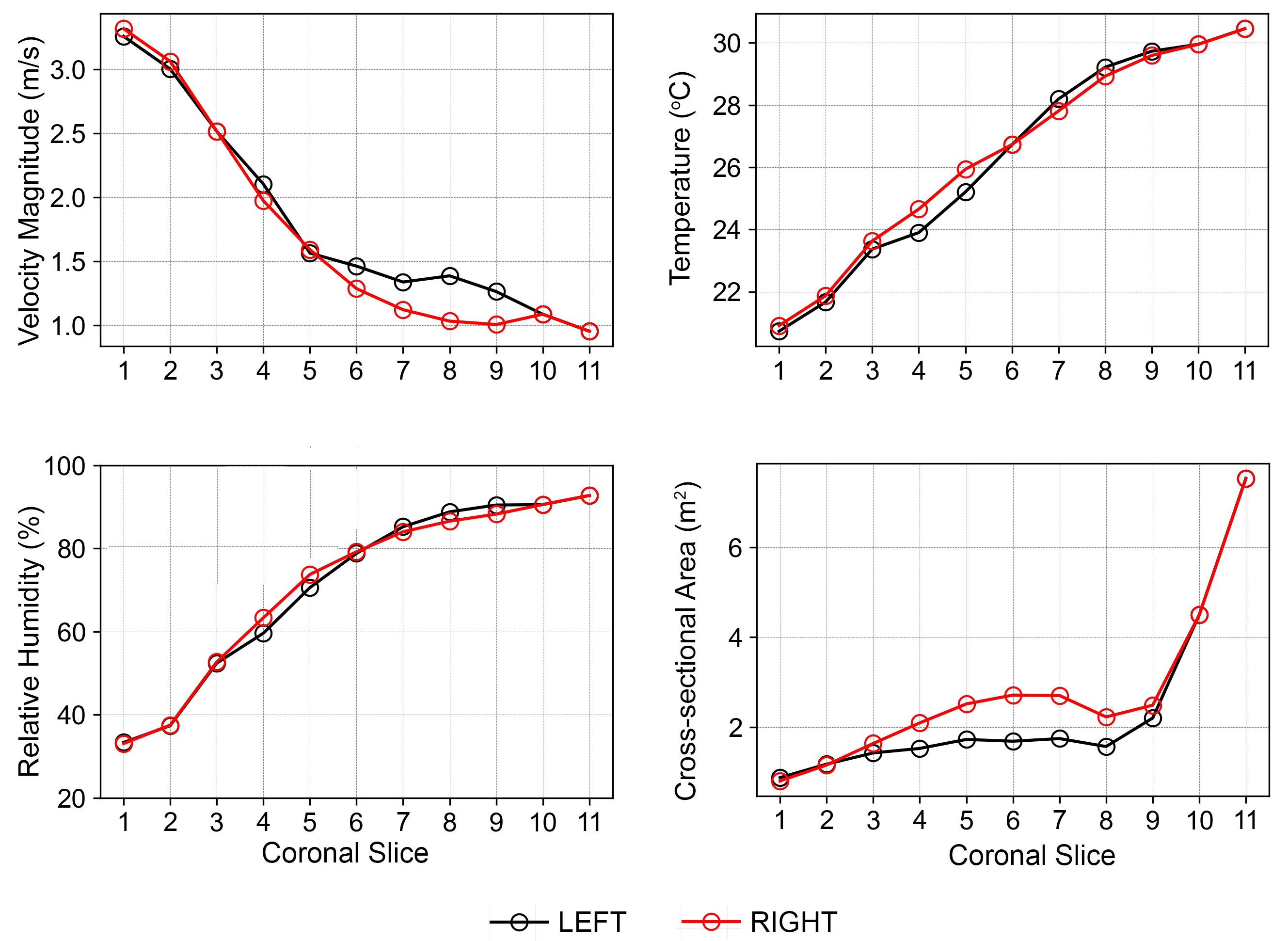}
	\caption{Averaged values of velocity, temperature, relative humidity, and cross-sectional area at each coronal slice for the No-Adhesion model.}
	\label{fig:noadh}
\end{figure}

\clearpage
\begin{figure}
	\centering
	\includegraphics[width=0.75\linewidth]{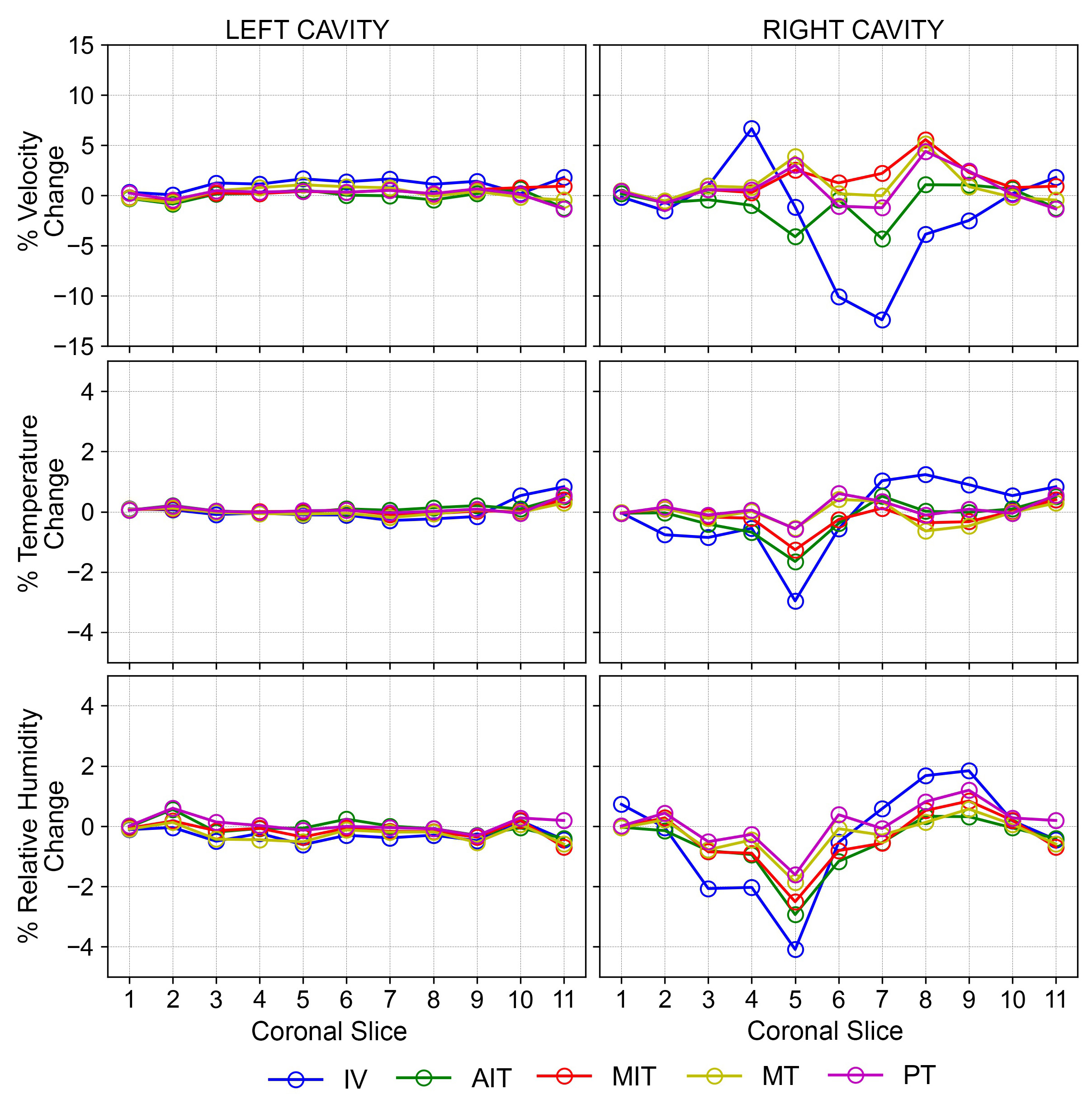}
	\caption{Velocity, temperature, and relative humidity difference between the five adhesion models and the baseline No-Adhesion model for the left and right cavities.}
	\label{fig:cont-difference}
\end{figure}

\clearpage
\begin{figure}
	\centering
	\includegraphics[width=0.85\linewidth]{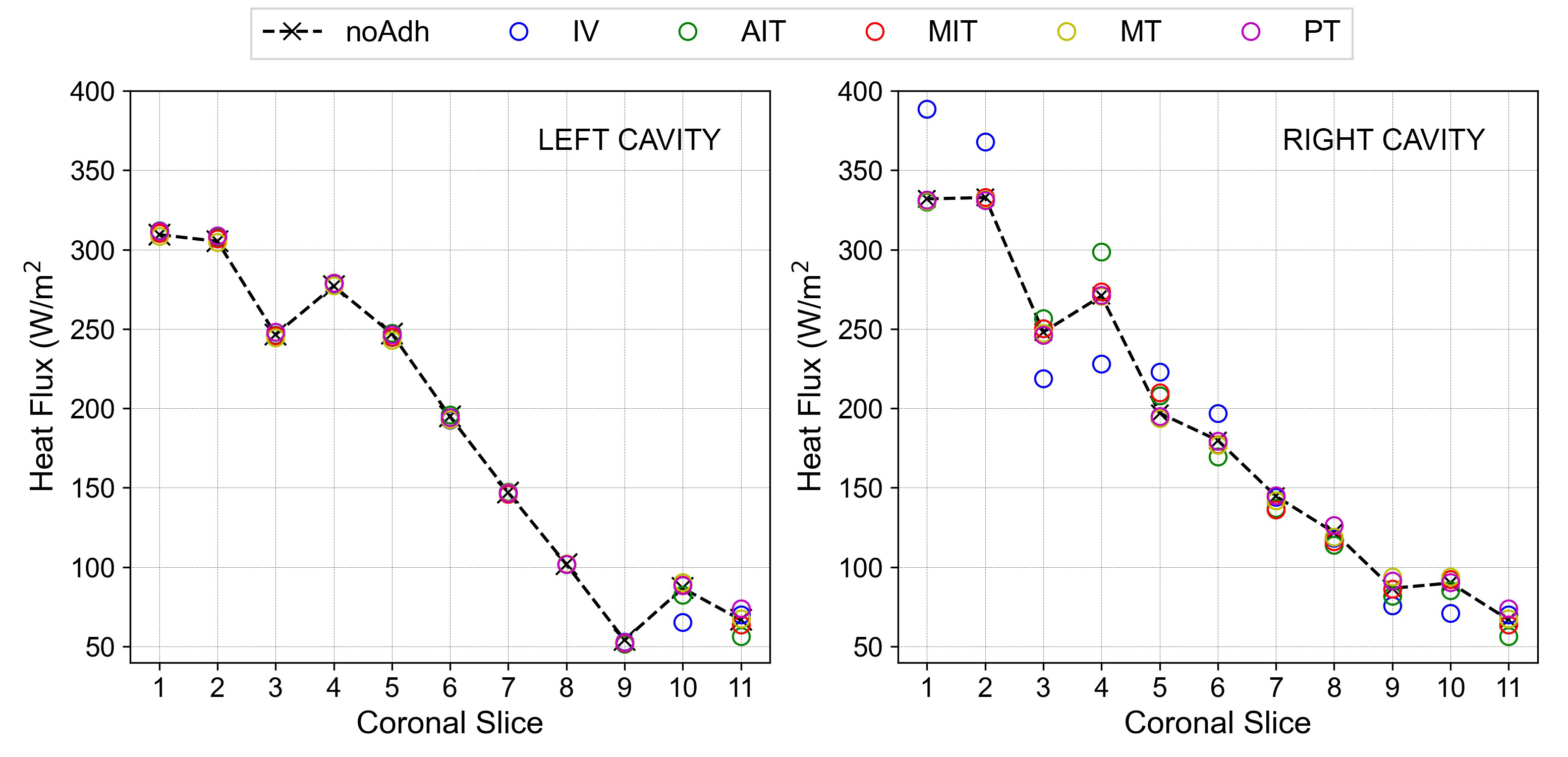}
	\caption{Heat flux at coronal slices for all models for the left and right cavities.}
	\label{fig:heatFlux}
\end{figure}

\clearpage
\begin{figure}
	\centering
	\includegraphics[width=.7\linewidth]{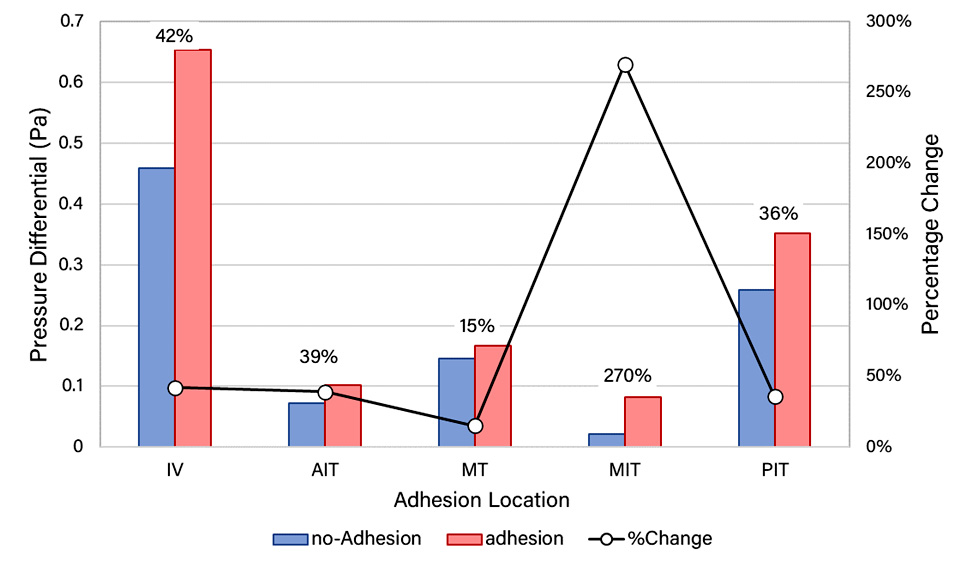}
	\caption{Comparison of the pressure drop across the adhesion location between the adhesion model and the no-Adhesion control model. The left vertical axis corresponds to the bar plots representing the pressure differential. The right vertical axis corresponds to the percentage change between the adhesion model and the no-Adhesion baseline model.}
	\label{fig:press}
\end{figure}

\clearpage

\begin{figure}
	\begin{subfigure}[b]{1\textwidth}
		\centering
		\includegraphics[width=0.67\linewidth]{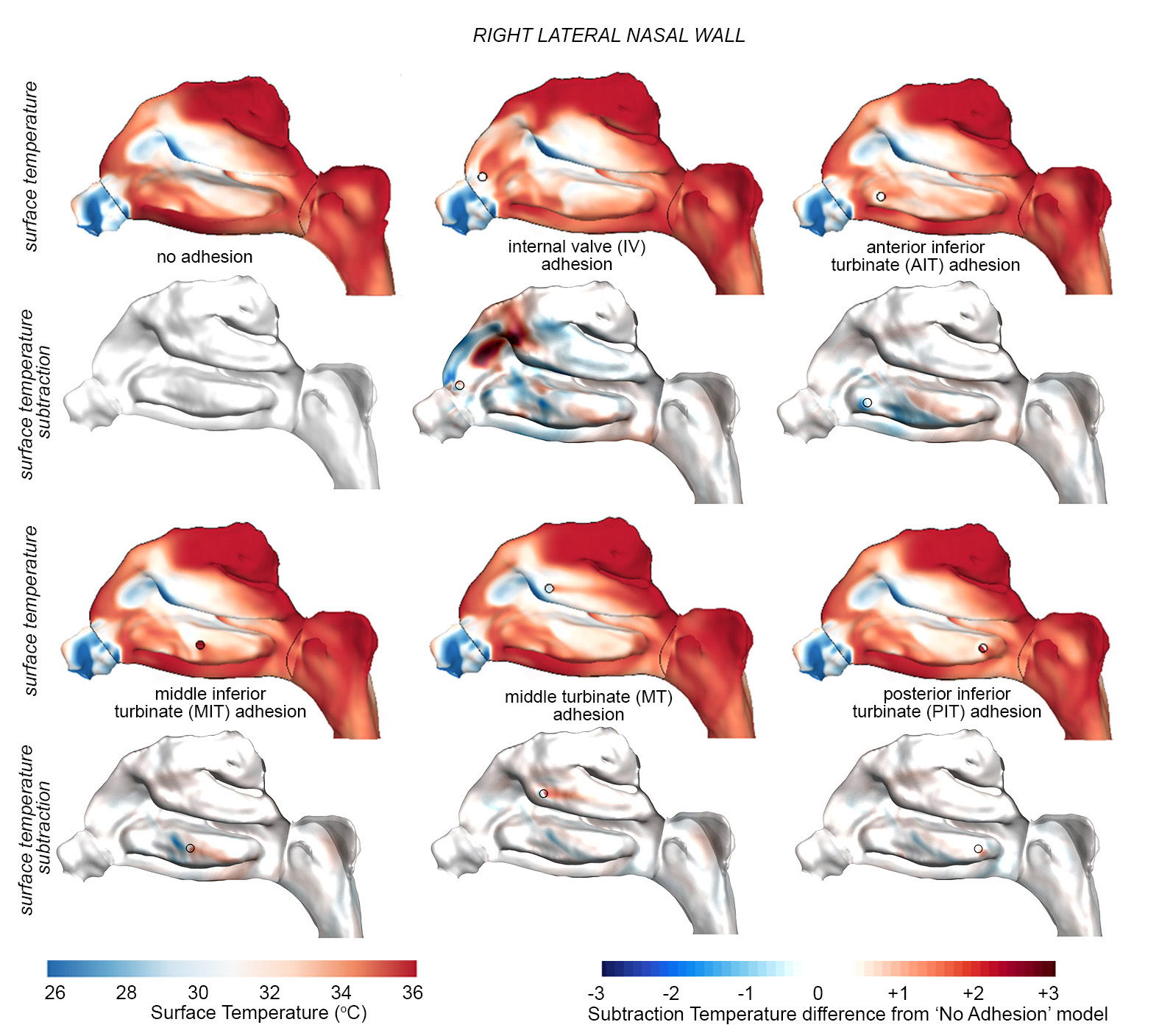}
		\caption{View of the right lateral nasal wall}
	\end{subfigure}
	\centering
	\begin{subfigure}[b]{1\textwidth}
		\centering
		\includegraphics[width=0.67\linewidth]{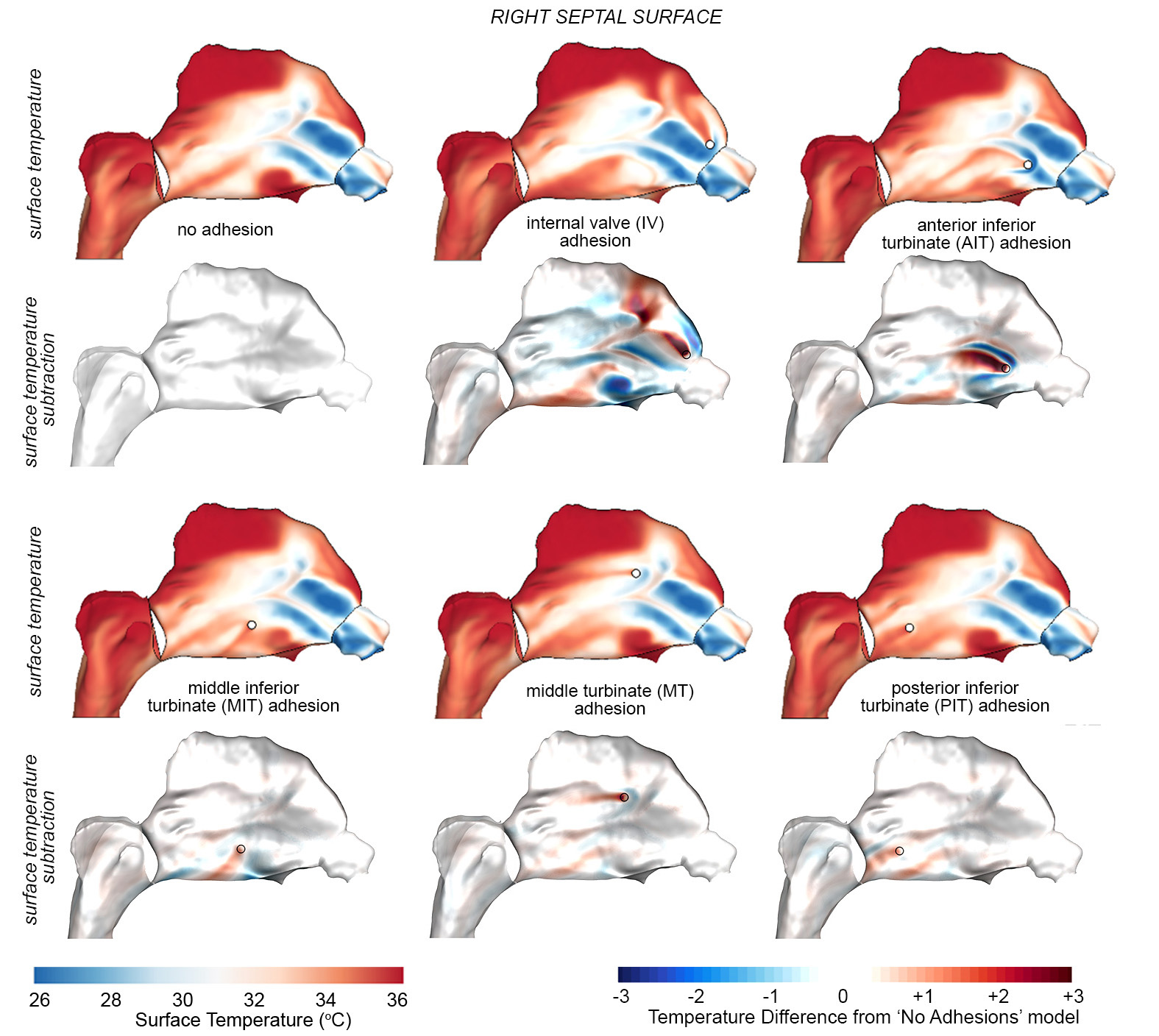}
		\caption{View of the right septal wall surface}
	\end{subfigure}
	\caption{Mucosal surface temperature for the right nasal airway.NA are shown as black circles. NA at; IV, interval valve; AIT, anterior inferior turbinate; MIT, middle inferior turbinate; PIT, posterior inferior turbinate MT, middle turbinate.}
	\label{fig:surfSubt}
\end{figure}

\clearpage

\section*{Supplementary Material}
\begin{figure}[h!]
	\centering
	\includegraphics[width=0.75\linewidth]{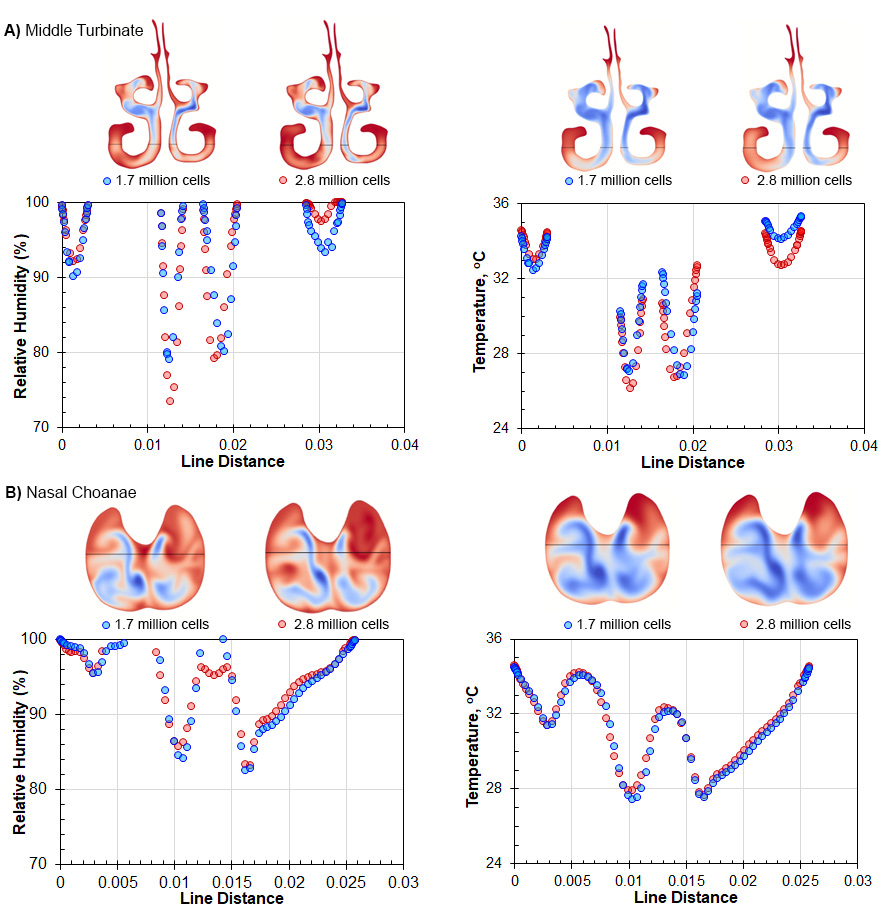}
	\caption*{\textbf{Supplementary 1}: Mesh independence comparison between the final mesh and a further refined model depicting the convergence of temperature and humidity parameters at a) Middle turbinate  and b) Nasal choanae coronal cross-sections}
\end{figure}

\clearpage
\begin{figure}[h!]
	\centering
	\includegraphics[width=0.5\linewidth]{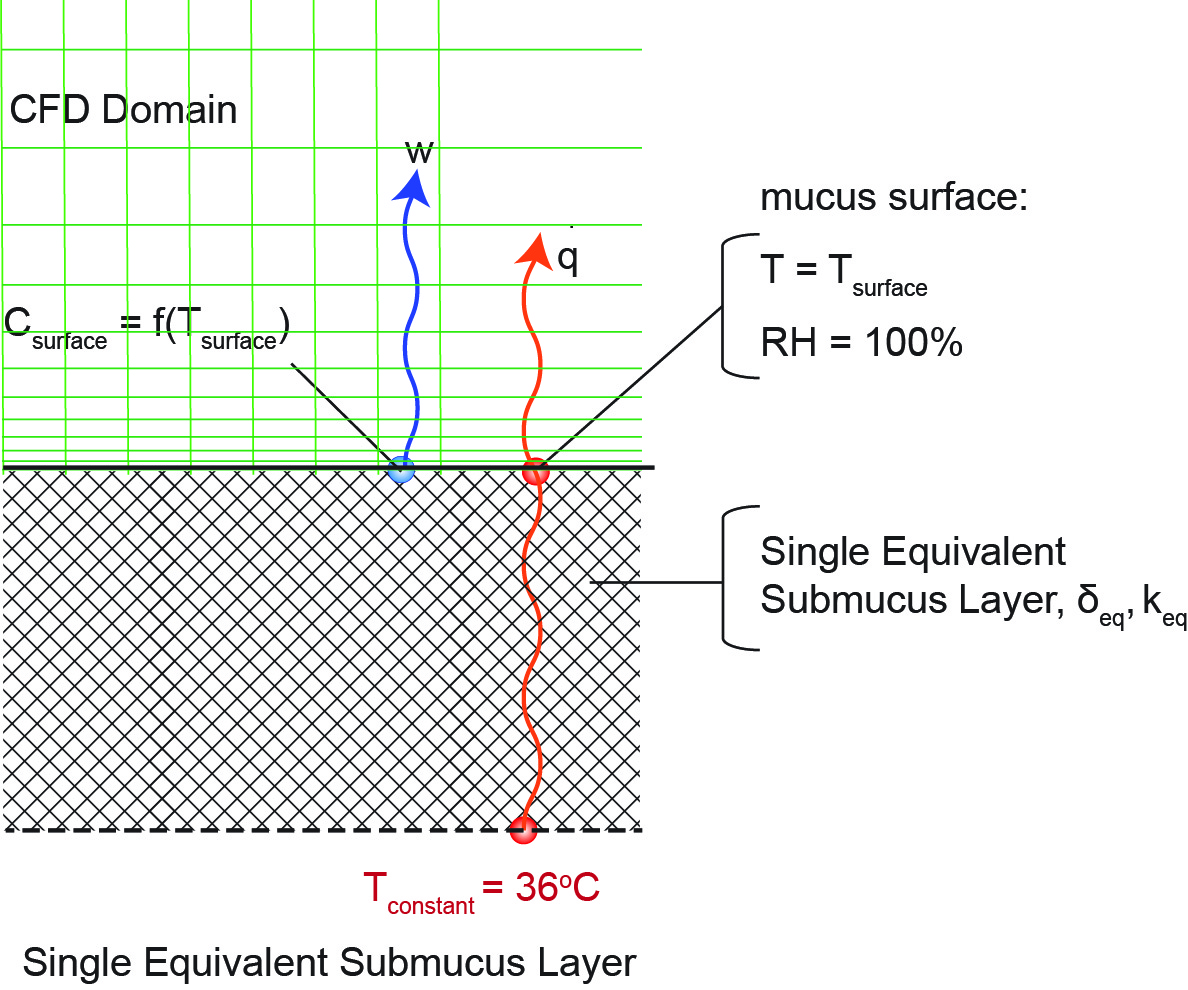}
	\caption*{\textbf{Supplementary 2}: Submucosal wall model for air conditioning}
	\label{fig:submucus}
\end{figure}

\clearpage
\begin{figure}[h!]
	\centering
	\includegraphics[width=1\linewidth]{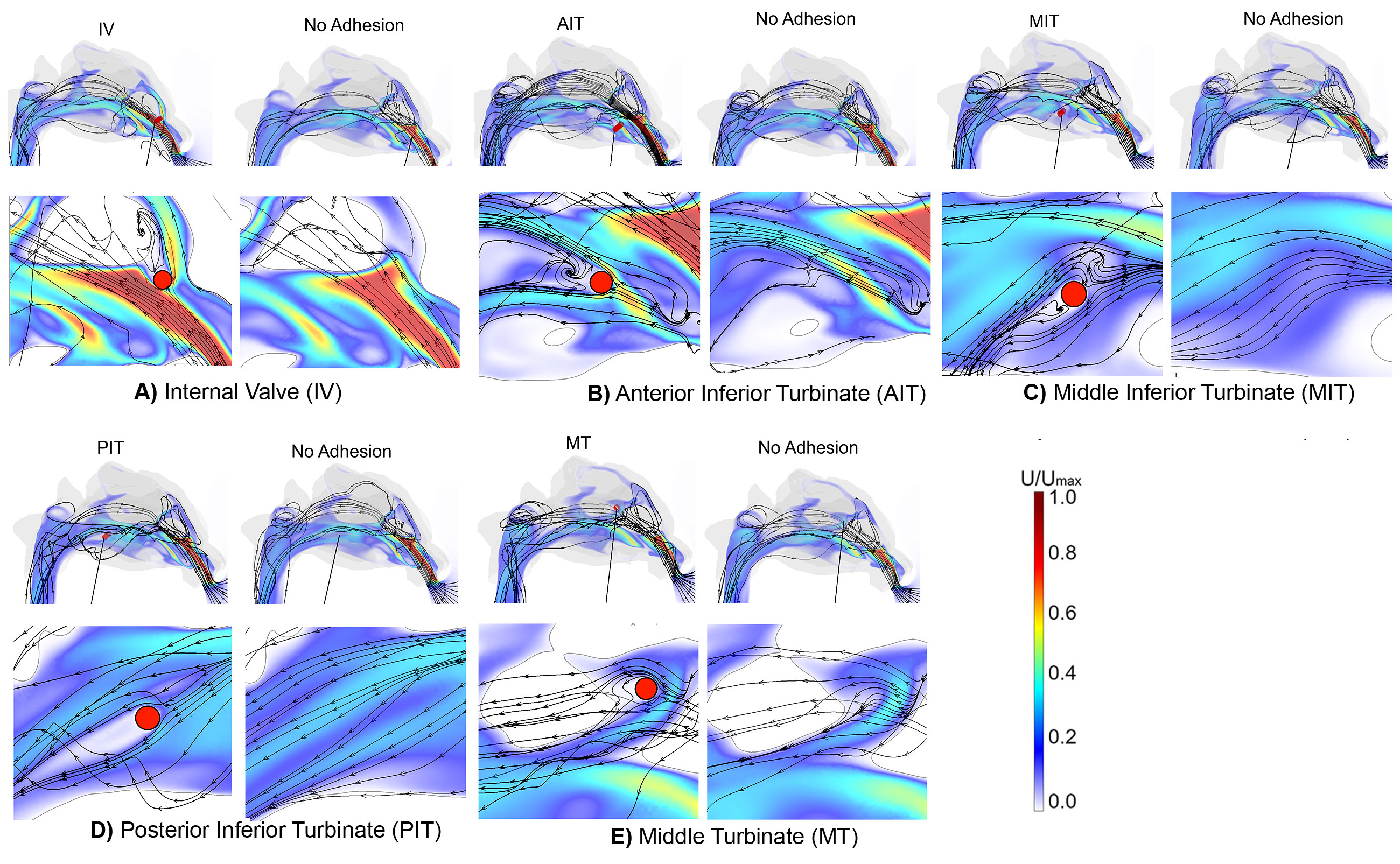}
	\caption*{\textbf{Supplementary 3}: Magnified lateral view of airflow streamline distribution in normal models and models with nasal adhesions (NA) located at (A) internal valve, (B) anterior inferior turbinate, (C) middle inferior turbinate, (D) posterior inferior turbinate and (E) middle turbinate. The semi-transparent red cylinders in each model indicate the position of the NA, whilst magnified views of the nasal cavity in the region of NA are shown below for each model (A-E). The IV model showed the main flow passing inferiorly and a smaller component superiorly, but both occurred at high velocity. In the downstream wake of the adhesion, recirculating flow was found. These represent swirling reversal of the airflow due to flow separation. Vortices in the wake of nasal adhesions are evident in all models, most notably in the IV and AIT models where the upstream laminar flow is at a higher velocity, compared to the other models.}
		\label{fig:streams-cont-v2}
	\end{figure}

\clearpage
\renewcommand{\arraystretch}{1.5}
\begin{table}[h!]
	\centering
	\caption*{\textbf{Supplementary 4}: Average surface heat flux at different regions of the nasal cavity for different models}
	\label{tab:heatflux}
	\resizebox{\textwidth}{!}{%
		\begin{tabular}{lllllll}
			\\ \hline & No adhesion (W/m$^2$)	& IV (W/m$^2$) &	AIT (W/m$^2$)  & MIT (W/m$^2$)  &	MT (W/m$^2$)  &	PIT (W/m$^2$)  \\ \hline 
			Right vestibule &	421.7 &	416.5 &	420.5 &	421.4 &	422.3 &	420.7 \\ \hline 
			Left vestibule &	378.6 &	383.7 &	379.9 &	379.9 &	381.8 &	378.2 \\ \hline 
			Right nasal valve &	161.1 &	233.6 &	168.5 &	160.3 &	160.4 &	159.9 \\ \hline
			Left nasal valve &	183.7 &	183.5 &	181.7 &	183.7 &	181.3 &	185.3 \\ \hline
			Right septum &	186.2 &	193.1 &	189.6 &	191.1 &	183.2 &	184.0 \\ \hline
			Left septum	& 184.6 &	183.2 &	184.9 &	183.1 &	183.2 &	184.8 \\ \hline
			Right lateral nasal wall &	129.5 &	113.0 &	141.8 &	125.2 &	130.2 &	131.6 \\ \hline
			Left lateral nasal wall &	142.4 &	142.1 &	123.8 &	141.6 &	141.4 &	142.0 \\ \hline
			Right inferior turbinate &	160.2 &	163.5 &	153.5 &	152.6 &	161.9 &	161.4 \\ \hline
			Left inferior turbinate &	199.4 &	199.2 &	197.3 &	198.0 &	198.0 &	198.2 \\ \hline
			Right middle turbinate &	263.0 &	248.9 &	241.6 &	240.3 &	232.9 &	241.1 \\ \hline
			Left middle turbinate &	210.3 &	208.7 &	213.2 &	209.5 &	209.4 &	212.1 \\ \hline
			Right superior turbinate &	2.2 &	15.6 &	2.1 &	2.0 &	1.4 &	2.0 \\ \hline
			Left superior turbinate &	0.6 &	0.5 &	0.7 &	0.6 &	0.5 &	0.7 \\ \hline
			Adhesion &	NA &	403.0 &	372.2 &	251.8 &	311.2 &	237.4 \\ \hline
			Nasopharynx	& 87.6 & 80.7 & 81.8 & 87.3 & 88.7 & 90.4 \\ \hline
			Right nasal passage &	164.5 &	167.4 &	163.3 &	162.9 &	163.9 &	165.1 \\ \hline
			Left nasal passage &	161.4 &	161.6 &	161.4 &	161.0 &	160.9 &	161.6 \\ \hline
			
		\end{tabular}%
	}
\end{table}

\end{document}